\documentclass[10pt]{article}
\usepackage{amsmath}
\usepackage{amssymb}
\usepackage{graphicx}

\usepackage{lastpage}
\usepackage{fancyhdr}
\usepackage[colorlinks = true,
            linkcolor = black,
            urlcolor  = black,
            citecolor = black,
            anchorcolor = black]{hyperref}
\usepackage{multirow}
\usepackage{lscape}
\usepackage{enumerate}
\usepackage{wrapfig}
\usepackage{lastpage}
\usepackage{times}
\usepackage{enumitem}
\usepackage{epstopdf}
\usepackage{color}
\usepackage{ulem}
\usepackage{makecell}

\setdescription{leftmargin=7mm,labelindent=0in}

\begin{document}

\noindent {\Large \bf Management strategies in a SEIR-type model\\ of COVID 19 community spread}\\

\noindent {\it Anca R\v{a}dulescu\footnote{Corresponding author: Department of Mathematics, State University of New York at New Paltz, New Paltz, NY 12561. Email: {\it radulesa@newpaltz.edu}}, Mathematics, SUNY New Paltz}

\noindent {\it Cassandra Williams, Departments of Mathematics, SUNY New Paltz}

\noindent {\it Kieran Cavanagh, Departments of Mathematics and Electrical Engineering, SUNY New Paltz}\\

\begin{abstract}
\noindent The 2019 Novel Corona virus infection (COVID 19) is an ongoing public health emergency of international focus. Significant gaps persist in our knowledge of COVID 19 epidemiology, transmission dynamics, investigation tools and management, despite (or possibly because of) the fact that the outbreak is an unprecedented global threat. On the positive side, enough is currently known about the epidemic process to permit the construction of mathematical predictive models. In our work, we adapt a traditional SEIR epidemic model to the specific dynamic compartments and epidemic parameters of COVID 19, as it spreads in an age-heterogeneous community. We analyze management strategies of the epidemic course (as they were implemented through lockdown and reopening procedures in many of the US states and countries worldwide); however, to more clearly illustrate ideas, we focus on the example of a small scale college town community, with the timeline of control measures introduced in the state of New York. We generate predictions, and assess the efficiency of these control measures (closures, mobility restrictions, social distancing), in a sustainability context.
\end{abstract}

\section{Introduction}

The COVID 19 outbreak originated in December 2019, from a single focus in the Wuhan region (China) and over the course of less than three months had spread to every continent except Antarctica, affecting at this point (start of August 2020) 213 countries and territories, with more than 18,000,000 worldwide infections (more than 4,500,000 in the US alone) and over 690,000 fatalities (over 150,000 in the US, and growing). Depending on the age and immune system of each individual, clinical manifestations vary from mild to severe, to life threatening. Most critical complications (like severe pneumonia, acute cardiac injury, septic shock) have been reported to have highest prevalence in the $>80$ age bracket (with the mortality rate increasing from children, to young adults and adults, to the elderly). In the incipient stages of the epidemic, the public concern was curbed by the relatively low overall COVID 19 mortality rate (initially around 3.3\%), comparable with the rates of less threatening viral epidemics, such as the seasonal flu, rather than with those of notorious outbreaks like Ebola (~50\%), or SARS (~10\%). What first started to raise a strong reason for concern was the faster spread ($R_0 \sim 3$) via only mildly symptomatic cases. It became increasingly apparent over the following weeks of the outbreak that its control required international coordination. Over six months down the line -- after unprecedented global and local travel bans, functional shutdown of many branches of economic, educational and social life; after city and even country-wide mandated quarantines, self isolation and social distnacing -- the epidemic is far from subsiding, raising huge survival, economic and sustainability concerns. After having transcended in many countries a first epidemic wave (during which the immediate focus was on clinical survival and organizing the first response of the health care systems), the focus has been shifting towards long-term mitigation, and facing subsequent waves. An additional emerging concern concentrates around the plethora of potentially life long health effects related to survival after COVID 19 infection. This concern is slowly starting to contribute to the  shaping of the mitigation process.

Based on recent findings, it has been established that the current epidemic has a specific signature. First, it exhibits a {\it unique communicability timeline}. The SARS-COV-2 virus has a relatively long incubation period (defined as the length of time between the individual's exposure to the virus, and the presence of symptoms).  The average incubation period has been estimated to be approximately 6 days, with observed variations between 2 and 27 days~\cite{Incubation_period}). However, multiple studies have shown that people infect others before their own symptoms develop~\cite{luo2020modes,jang2020cluster,tong2020potential,yu2020familial}. In addition, the degree of infectiousness varies for an individual along this presymptomatic stage, with a high surge around 2-3 days before symptom onset, following an initial latent period, with lower communicability. Newer epidemiological research of COVID 19 has also revealed a significant percentage of asymptomatic carriers (broadly estimated between 6-41\%~\cite{byambasuren2020estimating}). Since they never develop symptoms, they are unaware that they are carrying the virus. Their infectiousness appears to be significantly lower than that of individuals who are in different stages of developing symptoms, but may still significantly contribute to the epidemic dynamics. Distinction between presymptomatic and asymptomatic individuals is therefore very important when developing public health strategies to control transmission, but it is also very difficult to make and requires careful tracking strategies. Presymptomatic and asymptomatic transmission, combined with the limited testing and tracking resources, hamper detection and contribute to the pandemic spread.

The second signature aspect of COVID 19 is that  it presents {\it significant age differences in symptom development and prognosis.} Children and young adults exposed to the virus can be contaminated as much as the more advanced age groups. However, their milder symptoms can pass undetected, and they can more easily act as carriers of the virus, effectively unrestrictedly spreading it to others. On the other hand, the elderly population is more likely to exhibit serious to critical symptoms post exposure. The mortality rates differ correspondingly between their age groups, as will be described in more detail in the upcoming sections. The original focus of medical care was aimed towards increasing survival, especially in the high risk elderly. With fatality rates apparently subsiding worldwide, the emerging concern extends to the potential long-term health consequences, primarily in the younger populations, upon infection and recovery from COVID 19.

Third, it has been established that COVID 19 {\it may not confer long-term immunity}, allowing people to get reinfected. This is important on a clinical scale, since patients with subsequent reinfections may have diminished chances of survival. On the larger scale, the potential for reinfection not only adds to the rate of the spread, but also question viability of social measures like herd immunity.

The reaction to the overwhelming pandemic consisted of a combination of containment and mitigation strategies (primarily based upon social isolation), aimed to compensate for the clinical unpreparedness, and diminish, or control the load on the overburdened health care systems by  ``flattening the infection curve.'' Gradually over a few weeks, as more countries and territories became affected, global travel was shut down, universities and schools were closed, followed by bars, restaurants and other entertainment venues, and finally by most churches and other religious or spiritual gatherings. The general population was asked to either quarantine or observe social distancing, depending upon suspicion of active symptoms, and upon the gravity of the local situation. These measures have been directly influenced in most places by (1) limitations in testing abilities; (2) the unpreparedness of the clinical field, of medical resources and suppliers to cater to an outbreak of this magnitude; (3) the difficulty in providing an immediate prevention plan.

As the first infection peak waned in some countries and regions, the immediate concerns moved towards reopening strategies. The epidemiological factors weighing on reopening are tightly intertwined with economic, health and  societal needs (which press towards reopening more venues), and with the psychological impact of the long-lasting pandemic on individuals (one result of which is a wide-spread reluctance to continue abiding by social isolation and distancing rules). When planning for the return to ``normal,'' the same mitigation aspects that represent central discussion points are: (1) which destinations are safer to visit; (2) what aspects of social distancing are most effective, and (3) for how long should they be enforced. Over six months of world-wide clinical research on COVID 19 clinical and epidemiological aspects has lead to a better understanding of both viral and social dynamics, better classifying destinations by infection risk. For example, outdoor gatherings have been shown to pose significantly lower risk than indoor similar encounters; a visit the doctor's office is a considered a very low risk, while school attendance, or working in an office are considered a moderate risk, and going to the gym, or eating at an indoor bar are deemed as high risk behaviors~\cite{risklist}. The many research efforts converging on COVID epidemiology resulted on more extensive knowledge of how isolation, social distancing and hygiene contribute in combination to keeping infection at bay. After mixed directives at the start of the pandemic, the use of masks is now considered a crucial aspect in controlling the epidemic spread, especially in closed spaces, and in conjunction with respecting a minimum distance of six feet between individuals.

While our increasing knowledge from clinical observations and epidemiological studies has been shaping better mitigation strategies, answers to some of the critical questions cannot be extrapolated directly from field observations, and benefit from constructing testable mathematical models that can generate predictions. These can be used to inform our potential future directions, based on the current state of the epidemic, and also based on our better understanding of our actions thus far. A plethora of epidemic models of COVID 19 have already aimed at addressing some of these aspects~\cite{Wuhan,giordano2020modelling,carcione2020simulation,ferretti2020quantifying}. With our model, we will focus on a few specific questions.

A first important question regards the timeline of the original travel bans and isolation measures, and whether the timing, size (local versus global) and priorities were optimally weighted. The first institutions to close in most places were the schools and university campuses. Only then came retail and entertainment (such as gyms, restaurants and theaters). Essential services such as doctor's offices, food stores and pharmacies remained open to the general public, as well as many outdoor parks. Some of the last to be shut down were religious/spiritual services and gatherings; while supporting people psychologically or spiritually, these may have considerably added to the infection rate. A model can assist with establishing the impact of different shutdown combinations, by re-creating the conditions at time zero of the outbreak, and simulating how different strategies would have changed the outcome that we now observe. The answer to this question is of increasing importance, as hotspots are emerging in other areas across the US and in some places in Europe, and as we are beginning to see the threat of subsequent waves. Retrospective modeling may help public health and governmental officials learn from the first wave so that future waves are handled differently, resulting in a less detrimental impact.

The second question regards timing and efficiency of the social distancing requirements. It has been shown repeatedly that hygiene measures, in conjunction with respecting a six foot distance and wearing a mask work effectively towards diminishing the risk of infection. However, overwhelming evidence from media reports in many countries suggests that there is very wide variability in the population response to these measures (which may be mandated or not, depending on location). While in some regions social distancing was efficiently applied as a successful mitigation measure, in some of the regions which have been most affected, such measures are still being observed by only a relatively small percentage of the population. A model can illustrate the relative importance of social distancing versus decreasing actual mobility. Indeed, the general expectation is that a sufficiently long and efficient isolation will successfully curb the outbreak. However, that is not realistic. Not only are there necessities (e.g., for food, medical care or essential work) that constantly send people out of their homes in the path of exposure, but one also has to account for the psychological reaction to the long-term isolation measures, and for the increasing psychological tolerance to the epidemic (both feeding into increasingly risky behaviors). A model can incorporate all these aspects, and estimate the importance of social distancing measures not only under isolation, but as mobility is gradually being restored.

Finally, one obvious question that needs to be considered regards the timeline of reopening strategies. Reopening has been posing an extremely difficult optimization problem over the past few months, weighing on one hand the high risk of further epidemic spread, loss of human lives, and an enormous economic pressure on the heath care systems, and on the other hand sustainability issues related to exhaustion of resources and global economic shutdown. The difficulty of the problem is increased by the fact that the two parts to be reconciled are measurable in different units (loss of human life versus livelihood), and are tightly intertwined. To make matters worse, it is becoming increasingly apparent that, beyond its immediate life-threatening risk (which seems to be currently decreasing on an epidemiological scale), infection with COVID 19 may increase subsequent risk to potentially critical life-long health problems. Current optimistic estimates based on a preliminary vaccine quote months of necessary testing, and do not make it clear how administration of a vaccine would impact an infected, but asymptomatic individual. Modeling could generate informed predictions as to the most sustainable long-term mitigation strategies, including which venues are safest to reopen, and whether social distancing alone may offer sufficient protection while relaxing the mobility restrictions.

In this paper, we aim to focus on these basic questions, using a model specifically tailored to incorporate the signature of the COVID 19 dynamics, and the limitations in our response to it. Using a traditional SEIR setup that accounts for long incubation, different age compartments, asymptomatic and presymptomatic carriers, potential lack of immunity and minimal testing, we simulate the epidemic dynamics first within one community with a specific social pattern. We set out to understand the effects of the social measures that were imposed in this community upon the first epidemic surge. We do so based on the example of the response timeline in New York State; this can be, however, easily modified to match other states' timeline, or the widely different response strategies implemented in other countries. We follow up by exploring the need and efficiency of maintaining or relaxing such measures in the process of reopening, and as a long-term strategy.

\section{Methods}
\label{methods}

\subsection{Improving on the basic SEIR model}

One of the most traditional and relatively simple mathematical frameworks to study epidemics at the population level is the Susceptible / Exposed / Infectious / Recovered (SEIR) compartmental model. A classical SEIR model considers four compartments: the susceptible population $S(t)$ at time $t$ (i.e., healthy individuals who have not been exposed to the disease); the exposed population $E(t)$ (individuals who have contracted the virus but are not yet symptomatic); the infected population $I(t)$ (exhibiting signs and symptoms of the illness); the recovered population $R(t)$ (in an oversimplified view, the number of individuals who can no longer infect others). In a closed system which does not account for births or deaths, the sum of these compartments $N = S(t)+E(t) +I(t)+R(t)$ remains constant in time. The coupled dynamics of these compartments are described by the following system of equations:

\begin{eqnarray}
\frac{dS}{dt} &=& - \beta S(I+qE)/N \nonumber \\
\frac{dE}{dt} &=& \beta S(I+qE)/N - E/\delta \nonumber \\
\frac{dI}{dt} &=& E/\delta - I/\gamma \nonumber \\
\frac{dR}{dt} &=& I/\gamma
\label{SEIR_sys}
\end{eqnarray}

\noindent The parameter $\beta$ is the average number of contacts per person per time, multiplied by the probability of disease transmission via a contact between a susceptible individual, and an individual carrying the virus. The carrier can be either infected on exposed, with the fraction $SI/N^2$ representing the likelihood of an arbitrary contact to be between a susceptible and an infectious individual, and the fraction $SE/N^2$ corresponding to the likelihood of a contact to be between a susceptible and an exposed individual. The model allows the possibility that a contact with an exposed individual may have different probability of transmission than that made with an infected individual, which is reflected in the scaling factor $q$). The transition rate at which people are exposed then takes the form $-d(S/N)/dt = \beta S(I+qE)/N^2$, leading to the first equation $dS/dt = -\beta S(I+qE)/N$. The rate of transfer from the exposed to the infectious stage is a fraction $1/\delta$ of the number of exposed individuals, where $\delta$ is the average time for an exposed individual to become infectious. The rate of recovery is a fraction $1/\gamma$ of the infectious population, where $\gamma$ is the average time it takes a person to die or recover once in the infectious stage. This model has already been used in its original form for an early assessment of the epidemic in Wuhan, China~\cite{Wuhan}. We will adapt this model to the encompass recent epidemiological information in a few different ways.\\


\noindent \textbf{Transmission.} Clinical evidence confirms that COVID 19 transmission occurs from person to person through several different routes: contact with respiratory droplets generated by an infected individual through the breathing, sneezing, or coughing of an infected individual; direct (person-to-person) or indirect (hand-mediated) transfer of the virus from contaminated fomites to the mouth, nose, or eyes~\cite{transmission1}. The onset and duration of viral shedding and the period of infectiousness for COVID 19 are not yet known with certainty. Based on existing literature, the incubation period (the time from exposure to development of symptoms) of SARS-CoV-2 ranges typically from 2--14 days~\cite{infectious1}, with a mean of 5.2 days~\cite{Wuhan}. Based on current evidence, scientists believe that persons with mild to moderate COVID 19 may shed replication-competent SARS-CoV-2 for up to 10 days following symptom onset, while a small fraction of persons with severe COVID 19, including immunocompromised persons, may shed replication-competent virus for up to 20 days~\cite{infectious1}. With an $R_0$ value for COVID 19 estimated between 1.9--3.3~\cite{Wuhan,riccardo2020epidemiological,liu2020reproductive}, and an infectious period of 10--20 days~\cite{infectious1} (recovery rate $1/\gamma$ between 0.05 and 0.1), the infection rate $1/\delta$ was estimated between 0.07--0.5, and the transmission rate $\beta =R_0 \gamma$  between 0.1--0.3~\cite{Wuhan}. While traffic limiting measures act to diminish transmission by decreasing the overall number of contacts (preventing susceptible individuals and carriers from sharing the same space), social distancing measures such as disinfecting hands, observing physical distance, wearing masks, avoiding contamination by touching surfaces can all be seen as factors that act on decreasing the parameter $\beta$ (by diminishing the probability of disease transmission when a susceptible person and a carrier do share the same location).

In terms of viral load profile, SARS-CoV-2 peaks at around the time of symptom onset~\cite{subbarao2020sars,transmission2}, suggesting that the peak of the transmission may occur at an early stage of infection~\cite{transmission1}, likely even a few days prior to any detectable symptoms~\cite{presym1,wolfel2020virological,ferretti2020quantifying}. To capture these patterns in the transmission timeline, we further partition the traditional Exposed compartment, to differentiate between latent individuals $L(t)$ (who have been exposed to the virus, but are still in the latent, relatively noninfectious stage), and presymptomatic individuals $P(t)$ (who entered the high transmission stage, but are not yet symptomatic). \\


\noindent \textbf{Asymptomatic and presymptomatic transmission.} In our model, we account separately for presymptomatic and asymptomatic transmission, by introducing two new compartments, $P(t)$ and $A(t)$ respectively. To do so, we used recent epidemiological data that estimates the proportion of asymptomatic cases (the extent of truly asymptomatic infection in the community remains unknown). The proportion of people whose infection is asymptomatic likely varies with age due to the increasing prevalence of underlying conditions in older age groups~\cite{asym1}. Transmission from infected people without symptoms is difficult to study, since substantiating information needs to be gathered via detailed contact tracing~\cite{asym1}. Available data, mainly derived from epidemiological studies of cases and contacts, vary in quality and do not deliver consistent conclusions~\cite{transmission1}. In a recent review, the proportion of asymptomatic cases among positive diagnoses was estimated at 16\% (with a range from 6 to 41\%)~\cite{byambasuren2020estimating}. Another review, found 25\% of individuals to be asymptomatic at the time of the positive testing, but only 8.4\% of these remained so throughout the follow-up period~\cite{koh2020we}. Our corresponding model parameters were based on these estimated ranges.

In terms of infectiousness, it was suggested by data-validated modeling studies that pre-symptomatic transmission contributed to 48\% and 62\% of transmissions in Singapore and China, respectively~\cite{ganyani2020estimating}, which suggests a large transmission scaling factor  ($q$ value) corresponding to the pre-symptomatic compartment (shortly preceding symptom onset). On the other hand, the coefficient chosen for the Latent and the Asymptomatic compartment is relatively small, since the WHO stated in June that transmission from purely asymptomatic individuals is low~\cite{WHO}. However, the statement has been since then subject to assiduous controversy, and does not mean that asymptomatic transmission cannot have a considerable contribution to driving the growth of the COVID 19 pandemic, as other mathematical modeling studies have suggested~\cite{aguilar2020investigating,huang2020taking}. By means of our compartmental construction and realistic estimates of our epidemiological parameters, we include this possibility in our analysis.\\


\noindent \textbf{Mortality.} The high mortality rate in the COVID 19 pandemic requires that our model have a designated compartment for fatalities, which we will call $D$. The particular age dynamics and distribution of COVID 19 require age dependent mortality rates (as discussed below). The fatality compartment is the only compartment of the model with no further interaction with the rest of the epidemic system.\\

\noindent \textbf{Immunity.} Recent evidence suggests that a large percent of recovered individuals may develop immunity, but whether this is efficient in the long term remains questionable. Immune response to SARS-CoV-2 involves both cell-mediated immunity and antibody production. However, whether the detection of antibodies to SARS-CoV-2 indicates protective immunity has not yet been established~\cite{immun1}. Most individuals infected with SARS-CoV-2 display an antibody response between day 10 and day 21 after infection~\cite{zhao2020antibody,okba2020sars,liuevaluation,long2020antibody}; for modeling purposes, it can be therefore assumed that immunity is established around the same time as the cessation of viral shedding (which in our model we identify as ``recovery.'') SARS-CoV-2 antibody levels may remain over the course of seven weeks~\cite{xiao2020profile}, or at least in 80\% of the cases until day 49~\cite{zeng2020antibodies}, but results on the full longevity and efficiency of the antibody response are inconsistent. Primary infection with SARS-CoV-2 was shown to protect rhesus macaques from subsequent challenge and casts doubts~\cite{bao2020reinfection} on reports of re-infections with SARS-CoV-2~\cite{kellam2020dynamics}. Based on the observed dynamics of antibody waning in other (seasonal) coronaviruses, a study showed that the duration of protective immunity lasted 6-12 months~\cite{edridge2020human}. To reflect all this information, our model used different scenarios of immunity length from 60 to 365 days, when returning recovered individuals to the susceptible pool. Since the length of immunity is not a factor we are specifically investigating in this paper, we used for consistency a fixed immunity window of 180 days for all our illustrations. Longitudinal serological studies that follow patients' immunity over an extended period of time would be required to improve these estimates of the duration of immunity~\cite{ferguson2020report}.  \\


\noindent \textbf{Age differences in epidemiological profile.} To represent the wide age differences in epidemiological profiles of COVID 19, we introduce four age compartments: Children (0-18 years of age), Young adults (18-30), Adults (30-70) and Elderly ($>70$). The boundaries between these groups were set to in accordance to epidemiological data, but were adapted to also account for social mobility patterns (e.g., K-12 school versus college attendance). These four groups will have not only different infection/recovery/fatality parameters (reflecting different clinical profiles), but will also exhibit different social interactions (based on knowledge of age-based differences in social behaviors, and on how these were differentially altered in response to the outbreak).

Data from Germany shows that in symptomatic children, initial SARS-CoV-2 viral loads at diagnosis are comparable to those in adults~\cite{wolfel2020virological}, and that symptomatic children of all ages shed infectious virus in early acute illness~\cite{l2020shedding}. However, the percentages of post-exposure presence and severity of infection vary widely among different age groups. A review study that used age-specific case data from 32 settings in six countries quantified the differences in infection rates and symptom severity across ages, and estimated susceptibility and clinical fraction by age~\cite{nicholasage}. The study found age-varying susceptibility to infection by SARS-CoV-2, where children are less susceptible than adults to becoming infected on contact with an infectious person, and also experience no symptoms or subclinical (mild and unreported) symptoms on infection more frequently than adults. More specifically, 21\% (12–31\%) of infections in those aged 10 to 19 years resulted in clinical cases, which increased to 69\% (57–82\%) in adults aged over 70 years. The age-specific susceptibility profile suggested that relative susceptibility to infection was 0.40 (0.25–0.57) in those aged 0 to 9 years, compared with 0.88 (0.70–0.99) in those aged 60 to 69 years. We used this study to inform our model when choosing the asymptomatic proportions in each age compartment.

Long-term prognosis for COVID 19 patients also differs widely, with fatality rates increasing consistently with age~\cite{ferretti2020quantifying} (likely due to factors such as age-based variations in immune response, presence of co-morbidities, etc). Fatality rates recorded for the duration of COVID 19 have also varied in time, and with geographic region~\cite{age_death1} (e.g., between 1-6\% in the US, 6-15\% in Italy). When broken up by age, the worldwide data suggests a very small mortality rate of 0-0.2\% in children ages 0-18, slightly higher rates of 0.2-0.3\% in young adults 18-30 years of age, rates roughly between 0.3-3.6\% in adults 30-70 years, and significantly higher rates that go up to 6-20\% in the elderly individuals over 70 years of age~\cite{age_death1}. We used this data to inform our model when choosing the mortality versus recovery rates for each age compartment.

Finally, differences in contact and hygiene patterns among individuals of different ages can also be responsible for the differences in transmission, and subsequently in the number of infections within each age group~\cite{nicholasage}. The mobility patterns assumed by our model, together with their age-based differences, are described in detail in the following sections.


\subsection{Assembling age-compartmental dynamics}

Our basic compartmental model of propagation in a population of $N$ individuals will involve seven compartments $X \in \{S, L, A, P, I, R, D\}$. From the traditional model, we continue to use the original $SIR$ compartments. In addition, we introduce (as discussed above) four new compartments, to adapt to the specific characteristics of the COVID 19 epidemiology: $L$ represents the proportion of Latent individuals (who have contracted the virus but are not yet contagious); $A$ represents the compartment of Asymptomatic individuals (who have contracted the virus, may infect others, but will never present any symptoms); $P$ represents Presymptomatic individuals (who have contracted the virus, are able to infect others, and have not yet, but will soon develop symptoms); $D$ represents the fatalities compartment. Each age group $age \in \{ C,Y,A,E \}$ is therefore characterized by the compartments $S^{age}$, $L^{age}$,$A^{age}$,$P^{age}$, $I^{age}$, $R^{age}$, $D^{age}$. We can then write the following system, describing the epidemic coupling between compartments and age groups, and implements the provisions described above:

\begin{eqnarray}
\frac{dS^{age}}{dt} &=& - \beta(age) S^{age}(Q_I \Sigma_I + Q_A \Sigma_L + Q_P \Sigma_P+Q_A \Sigma_A)/N \nonumber \\
&+& (1-\rho) \left( [1-d(age)] \frac{I^{age}}{\gamma}  + \frac{A^{age}}{\theta} \right) + \frac{ R^{age}}{\varphi} \nonumber \\
\frac{dL^{age}}{dt} &=& \beta(age) S^{age}(Q_I \Sigma_I + Q_A \Sigma_L + Q_P\Sigma_P+Q_A\Sigma_A)/N - \frac{L^{age}}{\lambda_1} \nonumber \\
\frac{dA^{age}}{dt} &=& \alpha(age)\frac{L^{age}}{\lambda_1} - \frac{A^{age}}{\theta} \nonumber \\
\frac{dP^{age}}{dt} &=& [1-\alpha(age)]\frac{L^{age}}{\lambda_1} - \frac{P^{age}}{\lambda_2} \nonumber \\
\frac{dI^{age}}{dt} &=& \frac{P^{age}}{\lambda_2} - \frac{I^{age}}{\gamma} \nonumber \\
\frac{dR^{age}}{dt} &=& \rho\left( [1-d(age)] \frac{I^{age}}{\gamma} + \frac{A^{age}}{\theta} \right) - \frac{ R^{age}}{\varphi} \nonumber \\
\frac{dD^{age}}{dt} &=& d(age)\frac{I^{age}}{\gamma} 
\label{SEIR_ages}
\end{eqnarray}

\noindent The exposure rate $\beta$ is age dependent, based on age behavioral differences which may be riskier or more conservative (e.g., children touch their faces more often and may find social distancing more challenging than adults, young adults are more socially interactive than the elderly). In the next modeling steps, when people are allowed traffic to different locations, the rate $\beta$ will also be location dependent, since different places observe different hygiene protocols, and different social patterns. The second proportionality factor in the infection term changes to $Q_I \Sigma_I + Q_A \Sigma_L + Q_P\Sigma_P+Q_A\Sigma_A$, where $\Sigma_I = \sum_{age} I^{age}$, $\Sigma_L = \sum_{age} L^{age}$, $\Sigma_P = \sum_{age} P^{age}$, and  $\Sigma_A = \sum_{age} A^{age}$ . This is because individuals in a specific age group are exposed to infected, latent, asymptomatic and presymptomatic people from all four age groups that interact in one location.

Healthy individuals leave the $S$ compartment when exposed to the virus, and enter the latency compartment $L$, at a rate equal to the transmissions parameter $\beta$ multiplied by the ``exposure risk.'' The risk is calculated as a weighed sum of the proportion of spreaders in the population, with larger scaling factors assigned to the Presymptomatic and Infectious compartments ($Q_P$ and $Q_I$, respectively, and a lower scaling factor $Q_A$ given to the Latent and Asymptomatic compartments, according to the current literature on transmission dynamics). Latent individuals leave the compartment at a rate that reflects the number $\lambda_1$ of days estimated for the virus to start more pronounced shedding (corresponding to a stage of more dramatic transmission by a person who will subsequently develop symptoms). However, since not all individuals further develop symptoms, an age-dependent proportion $\alpha(age)$ of the Latent individuals will move into the Asymptomatic compartment, leaving the remaining proportion $1-\alpha(age)$ to join the high transmission Presymptomatic compartment.  Asymptomatic individuals will recover to the $R$ compartment without any further transitions, at a rate reflecting the time $\theta$ estimated for the viral shedding to subside. Meanwhile, Presymptomatic individuals go on to develop symptoms and enter the Infectious compartment, at a rate inverse proportional to the time $\lambda_2$ estimated for the presymptomatic stage. Symptomatic individuals leave the compartment at a rate based on the recovery time $\gamma$; they may recover (with potential immunity), or die. In our model, the age-specific parameter $d(age)$ accounts for the fatality percentage. A fraction $\rho$ of the survivors recover with limited immunity, hence they remain in the Recovered compartment, before re-joining the Susceptible compartment at a rate reflecting the approximate duration  of the immunity $\varphi$. Empirical parameters ranges, their corresponding sources and the values used in our model are listed in Table~\ref{param_values}.

\begin{table}[h!]
\begin{center}
{\renewcommand{\arraystretch}{1.25}
\begin{tabular}{|l|l|c|c|c|c|c|}
\hline
 {\bf Parameter} &   {\bf Description} & \multicolumn{4}{c|}{{\bf Model's Value}} & {\bf Source}  \\
 \cline{3-6}
  &  &   C & Y & A & E &  \\
 \hline
$\beta$  & Transmission rate at destinations & \multicolumn{4}{c|}{0.08-0.1 (See Table~\ref{parameters})} & \cite{Wuhan} \\
\hline
$Q_I$ & Transmission scaling factor for $I$ & \multicolumn{4}{c|}{1} &  ~\cite{ganyani2020estimating}\\
\hline
$Q_A$ & Transmission scaling factor for $A$ and $L$  & \multicolumn{4}{c|}{ 0.2} & *  \\
\hline
$Q_P$ & Transmission scaling factor for $P$ &  \multicolumn{4}{c|}{1.2} & ~\cite{ganyani2020estimating} \\
\hline
$\lambda_1 + \lambda_2$ & Incubation period &  \multicolumn{4}{c|}{6 days} &  ~\cite{infectious1,Wuhan}\\
\hline
$\lambda_1$ & Latency period & \multicolumn{4}{c|}{4 days} & ~\cite{presym1,wolfel2020virological,infectious1,Wuhan} \\
\hline
$\lambda_2$ & Presymptomatic period & \multicolumn{4}{c|}{2 days} & ~\cite{presym1,wolfel2020virological}\\
\hline
$\alpha$  & Asymptomatic incidence &  80\%  & 60\% & 40\%  & 20\% & \makecell{~\cite{byambasuren2020estimating,koh2020we}}\\
\hline
$\theta$ & Length of asymptomatic transmission &  \multicolumn{4}{c|}{12 days} & ~\cite{infectious1} \\
\hline
$\gamma$ & Time to recovery or death &  \multicolumn{4}{c|}{10 days} &  ~\cite{infectious1}   \\
\hline
$d$ & Fatality rate &  0.1\%  & 0.25\%  & 2\%  & 6\% & ~\cite{age_death1} \\
\hline
$\rho$ & Immunity incidence & \multicolumn{4}{c|}{0.8} & ~\cite{zeng2020antibodies} \\
\hline
$\varphi$ & Immunity length &  \multicolumn{4}{c|}{180 days} &  ~\cite{xiao2020profile,zeng2020antibodies,kellam2020dynamics,edridge2020human} \\
\hline
\end{tabular}}
\end{center}
\caption{\small {\it {\bf Summary of model parameters}, along with the corresponding references to the empirical ranges in existing literature.}}
\label{param_values}
\end{table}

\subsection{Incorporating social dynamics}

Finally, we expand the model to incorporate compartmental daily dynamics, allowing people age-specific mobility to different locations with potentially different exposure rates. To fix our ideas, we study a small ``college town'' community, including all age groups in equal proportion (1000 individuals), in which the contamination is initiated by two exposed adults. Daily travel is designed to reflect this profile, and our model is specifically tailored towards analyzing (1) the efficiency of epidemic mitigation measures, mandated to such communities as the outbreak was developing, and (2) the optimal timeline and conditions for reopening, which is one of the crucial open questions in all communities, as well as at a global stage. The model can be easily adapted to reflect a community with a different social profile. Our future work is focused on extending the model to multiple coupled communities, and aims to understand how the behavior patterns of one can easily affect the others.

In our current model, during each day, people travel from home to one of the following seven locations: medical office (e.g., doctor, hospital); shops (food, pharmacy); church (religious, spiritual gatherings); university campus (young adult education); school (children education); park (children entertainment); bars/restaurants (adult social and entertainment venues). Table~\ref{extended_table} specifies the baseline values of the exposure rate $\beta$ for each destination and age group, making educated estimates of the effects of hygiene restrictions and specific social interactions in each place.

To keep track of the day's dynamics to and from different locations, we use a mobility $7 \times 7 \times 4$ array $M$, so that each entry $M(place,X,age)$ specifies what fraction of the community travels to each location $place$, from each SEIR compartment $X$, for each age group $age$. These entries are time-dependent, allowing us to investigate not only extent, but also timing of quarantines and isolation measures.

For simplicity, in this first iteration of the model we assume that individuals may travel to at most one place each day, and spend there a specified amount of time, which is the same (6 hours) across places, epidemic compartments and age groups. The SEIR dynamics at each of the destinations is different, and different than the corresponding dynamics at home (where the exposure rate is assumed smaller than at all destinations).

\begin{table}[h!]
\begin{center}
{\renewcommand{\arraystretch}{1.25}
\begin{tabular}{|l|c|c|c|c|c|c|c|c|}
\hline
 {\bf Age group} & Home & Doctor & Store & Church & Campus & School & Park & Restaurant \\
\hline
 Children & $\beta_0$ & $\beta$ & 2$\beta$ &  2$\beta$ &  2$\beta$  &  2$\beta$ &  2$\beta$ & 3$\beta$\\
 Young &  $\beta_0$ &$\beta$  & 1.5$\beta$ & 1.5$\beta$ & 1.5$\beta$  & 1.5$\beta$ & $\beta$ & 3$\beta$\\
 Adults  &  $\beta_0$ &$\beta$  & 1.5$\beta$ & 1.5$\beta$ & 1.5$\beta$  & 1.5$\beta$ & $\beta$ & 3$\beta$\\
 Elder &   $\beta_0$ &$\beta$  & $\beta$ & $\beta$ & $\beta$  & $\beta$ & $\beta$ & 2$\beta$\\
\hline
\end{tabular}}
\end{center}
\caption{\small {\it {\bf Qualitative profile of exposure rates, for the community destinations.} The baseline value of $\beta$ was obtained as the product between the inverse of the time length $\theta$ during which the individual spreads the virus, and the reproduction value $R_0$. The reproduction value for this outbreak was estimated as $R_0 =3$~\cite{SEIR}. This led to $\beta=0.1$, which was lowered to $\beta_0=0.08$ in case of the home exposure rate, and adapted multiplicatively for each destination and age to reflect the corresponding patterns. For example: the likelihood of exposure for an adult is 3 times higher at a bar that at the doctor. In terms of age variability: children exhibit behaviors that make them more prone to exposure, the elderly are more careful. }}
\label{extended_table}
\label{parameters}
\end{table}

In our mathematical simulation, we dispatch theoretical people to all destinations in the morning, according to the day's mobility array (Appendix A, Tables~\ref{tableA1}-\ref{tableA4}), which accounts for the fraction of each compartment and age group in the community that chooses to visit that particular location during the specified day. To reflect the fact that limitations in testing and contact tracing may allow virtually unrestricted mobility in absence of specific symptoms, we assumed similar mobility of the $E$, $P$ and $A$ compartments to that of the $S$ compartment. Mobility from the Infectious compartment is drastically reduced (except traffic for health care), and subsequently restored after recovery. Also for simplicity, the mobility array is consistent throughout six days of the week. Every seventh day (Sunday), the primary activity is attendance of spiritual or social gatherings (broadly labeled in our destination vector as ``church''). We study the dynamics and effects of such periodic large gatherings separately, since they have been identified by field studies as a critical contribution to the overall epidemic dynamics.

At each destination, the coupled compartmental dynamics is applied for 6 hours. Compartmental dynamics with home parameters is applied to the home population, for an equal amount of time. Upon return from the destinations, home dynamics is applied for the rest of the day (18 hours). The code is implemented in the Matlab package, using an Euler method with step-size $h = 15$ minutes, tracking the progress of the outbreak within the community for 500 days~\cite{codes}, which is a rough estimate of the time needed for clinical prevention and treatment development. This observation window can be adapted to incorporate future changes to this clinical timeline.

Each destination $place$ has four age-specific exposure rate parameters $\beta(place,age)$, specified  for our model in Table~\ref{extended_table}. That is because the ability to maintain hygiene and social distance varies between places (e.g., doctor versus bar), and between ages (with children, for example, being less likely to abide by the strict rules of either hygiene or distancing). Once at home, the parameter rate assumes a smaller value $\beta_0$. One significant limitation of this approach is that it cannot capture specifics of the home dynamics, such as household structure (and the higher likelihood of infection of other household members from an infected person). This will be further discussed in the Limitations section, and will be the subject of future work.

We will study the effects of practicing social distancing at specific locations and in a residential setting, by varying the value of the respective parameter $\beta$. We will also study the effects of imposing early and late shutdowns of various destinations, by altering the mobility array. In this project, we investigate the schedule of measures that was government mandated in New York State communities (closing schools and campuses, then restaurants and bars, then churches, etc). We then investigate the effects of different opening timelines, and the differential consequences of relaxing the restrictions on mobility versus lightening the social distancing measures.

\section{Results}

\begin{figure}[h!]
\begin{center}
\includegraphics[width=\textwidth]{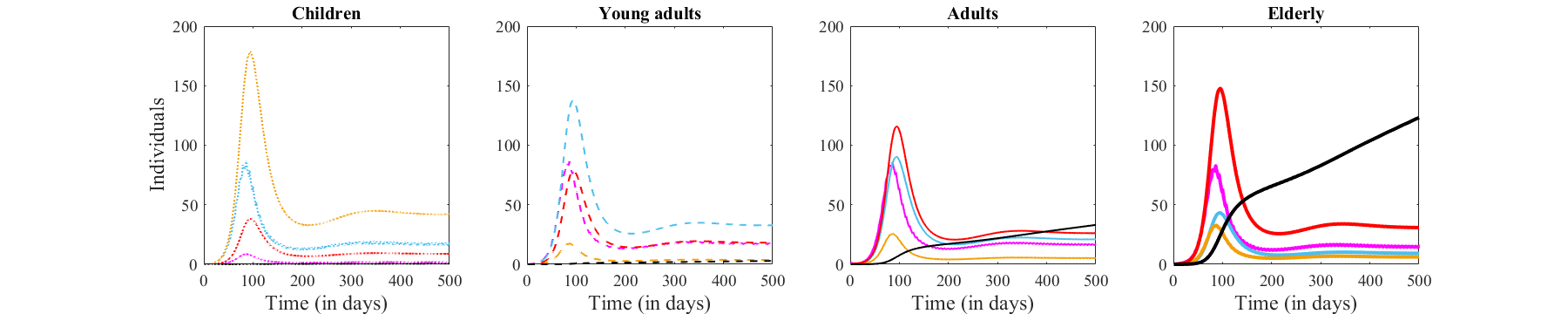}
\includegraphics[width=\textwidth]{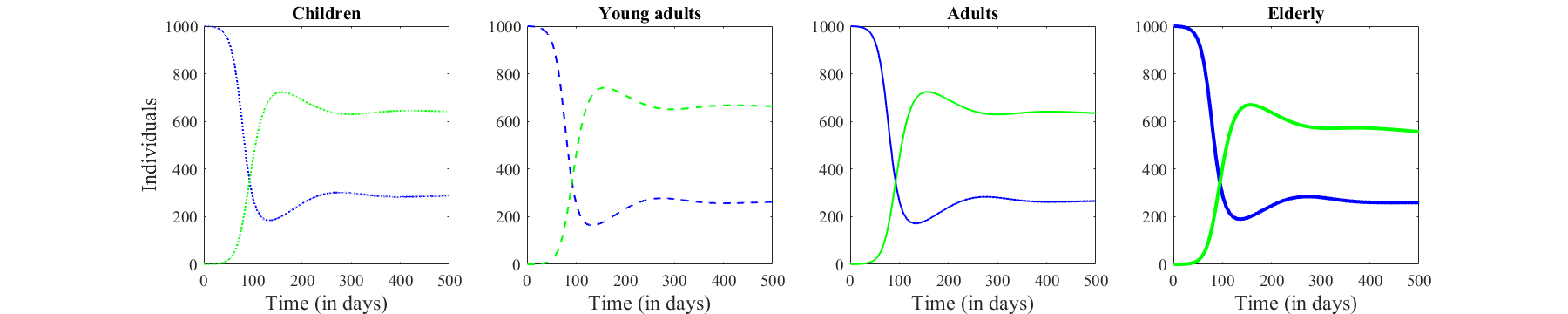}
\end{center}
\caption{\small {\it {\bf System dynamics in absence of any preventive measures.} Each column shows the evolution of one age group (from left to right: Children, Young adults, Adults and Elderly). The Symptomatic and Recovered compartments are shown in separate panels, for clarity of the illustration (larger scale than the other compartments). The number of Susceptible individuals is shown in blue, the Latent in pink, Asymptomatic in cyan, Presymptomatic in orange, Infected in red, the Recovered in green and the Fatalities in black.}}
\label{ages_dynamics}
\end{figure}

\begin{figure}[h!]
\begin{center}
\includegraphics[width=\textwidth]{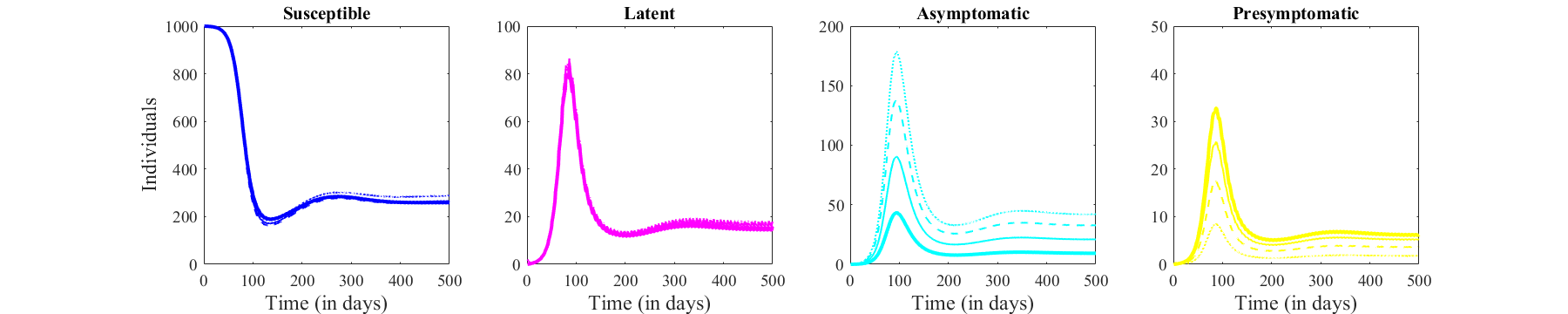}
\includegraphics[width=0.75\textwidth]{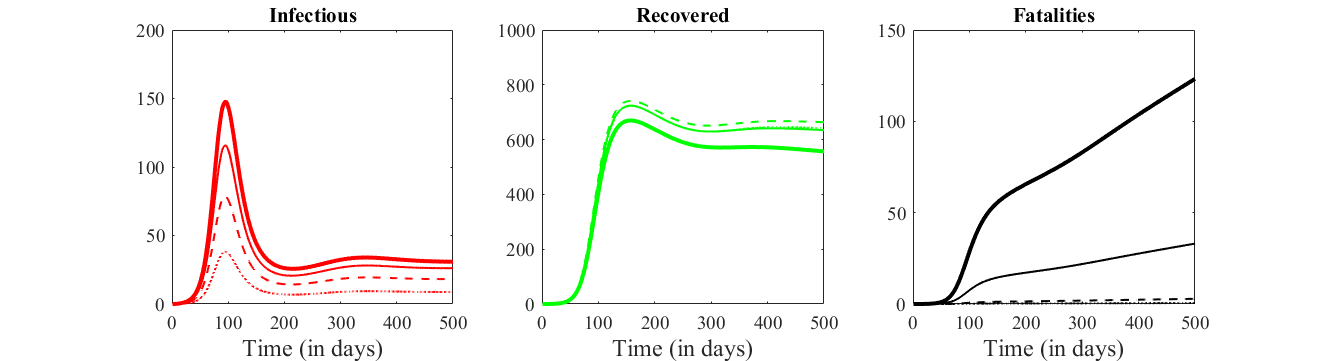}
\end{center}
\caption{\small {\it {\bf System dynamics in absence of any preventive measures.} Each panel shows the dynamics in one model compartment (from left to right: Susceptible, Exposed, Infected, Recovered and Dead.) In each panel, each age is represented by a different line type: Children as a dotted line, Young adults are a dashed line, Adults as a thin solid line and Elderly as a thick solid line.}}
\label{SEIR_dynamics}
\end{figure}

Figures~\ref{ages_dynamics} and~\ref{SEIR_dynamics} illustrate the long term dynamics of the system constructed in Section~\ref{methods}, in absence of any isolation, quarantine or closure measures. Figure~\ref{ages_dynamics} illustrates the interplay between model compartments (shown in different colors) for each of the four age groups (captured as a different panel). Figure~\ref{SEIR_dynamics} illustrates a comparison between age groups (shown in different line styles) in each of the seven compartments (shown in separate panels). There is a period of seemingly exponential growth in infections, followed by a peak at around 100 days, and a decline, with the number of fatalities monotonically increasing towards a different asymptotic value for each group (largest in the elderly, smaller in adults, etc). We will observe how these dynamics are affected by shutting down different locations at a specific time in the system's evolution. In further illustrations, we will focus in particular on the number of infected cases (relevant to the current stage of the epidemic, and to the strain on the health care system), on recovered individuals (who may be subject to life long consequences, whether they had been symptomatic of not), and on fatalities (as a traditional measure of epidemic impact in terms of life loss).

\begin{figure}[h!]
\begin{center}
\includegraphics[width=0.32\textwidth]{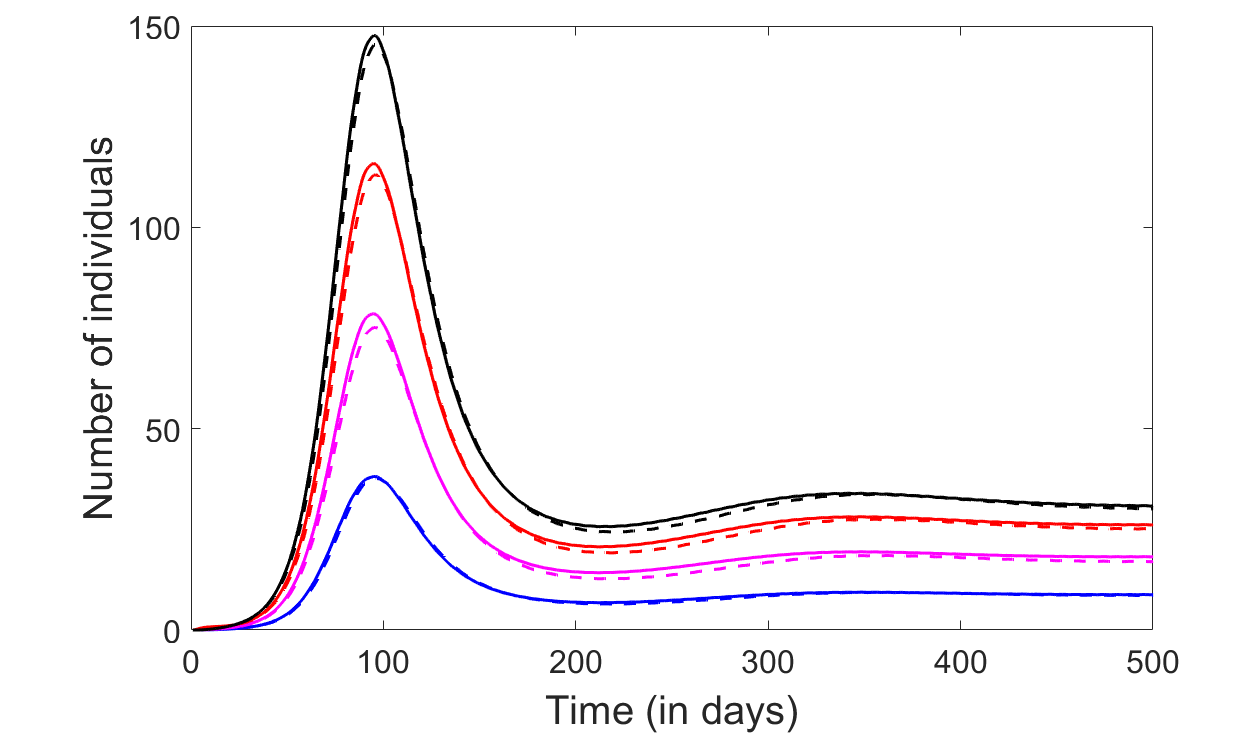}
\includegraphics[width=0.32\textwidth]{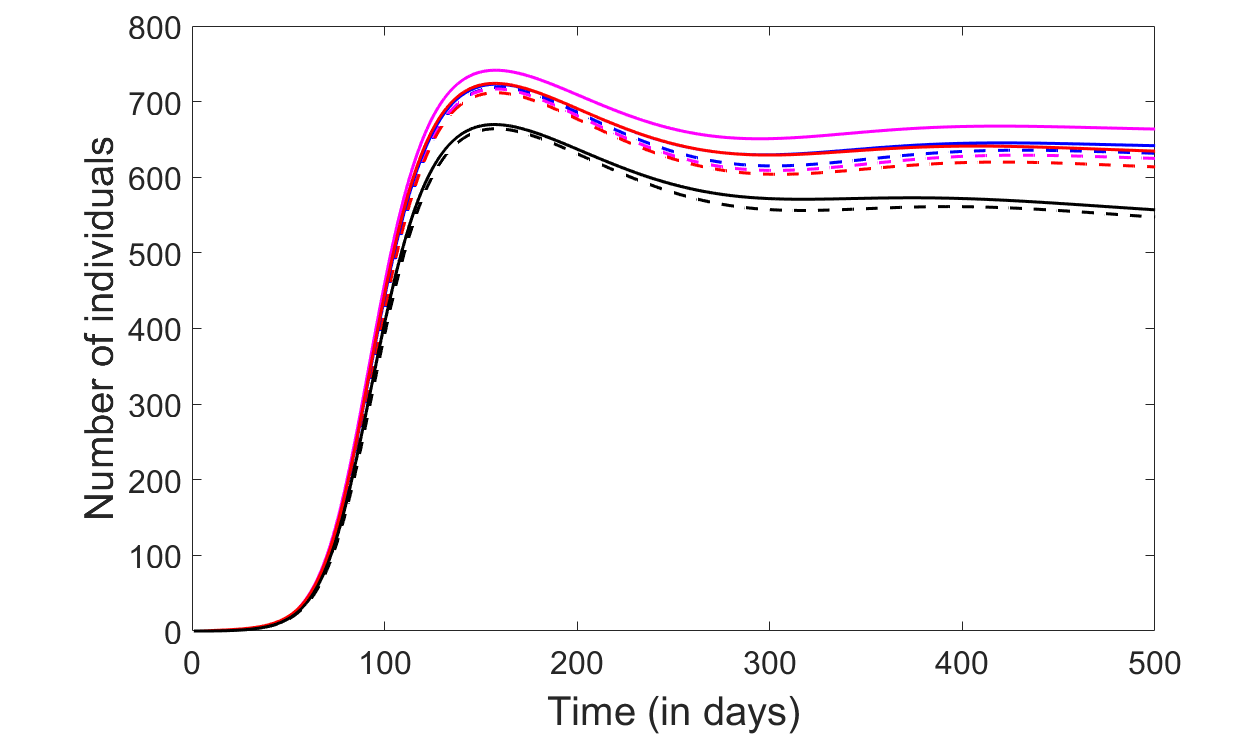}
\includegraphics[width=0.32\textwidth]{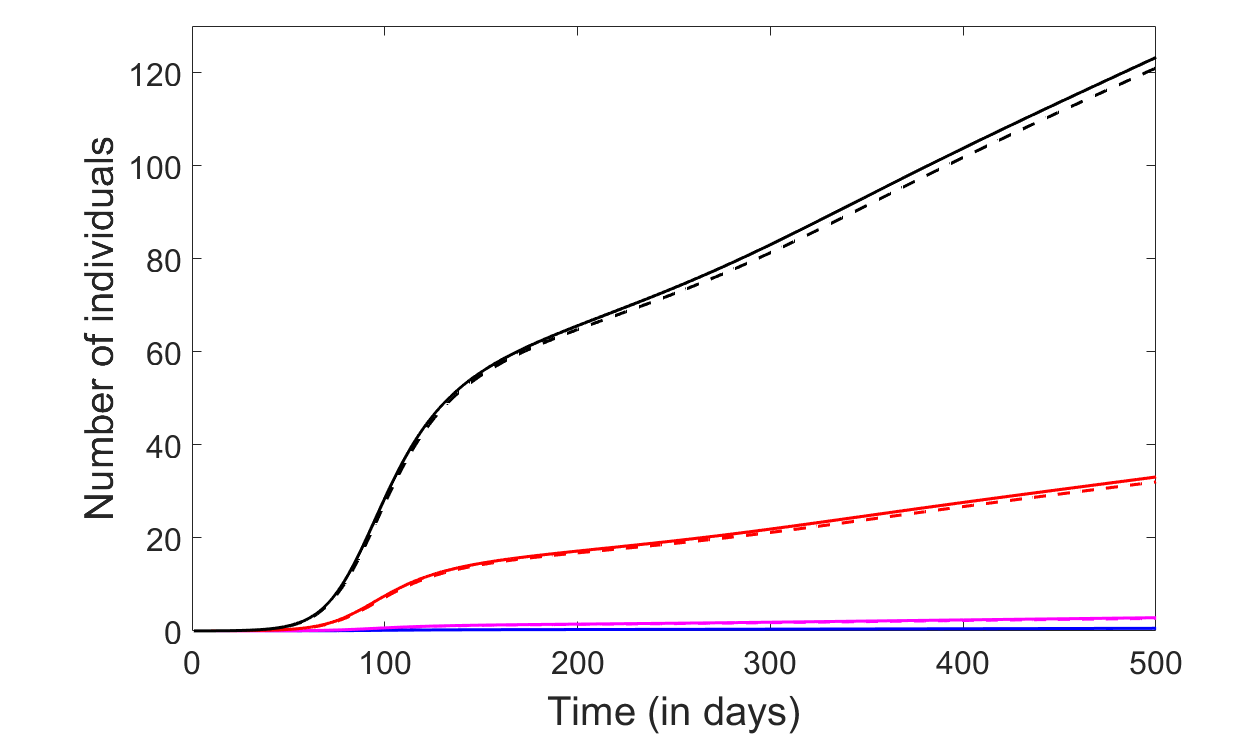}
\end{center}
\caption{\small {\it {\bf Effect of the campus closure on the systems dynamics.}  The left panel shows the rise and fall of the infected compartment, the center panel shows the recovered compartment and the right panel shows the accumulation of fatalities. Each age group is represented in one color: Children (blue), Young adults (pink), Adults (red) and Elderly (black). The solid curves illustrate the solutions in absence of closures; the dashed curves show the effect of the campus closure implemented 10 days after infection.}}
\label{campus_closes}
\end{figure}

\begin{figure}[h!]
\begin{center}
\includegraphics[width=0.32\textwidth]{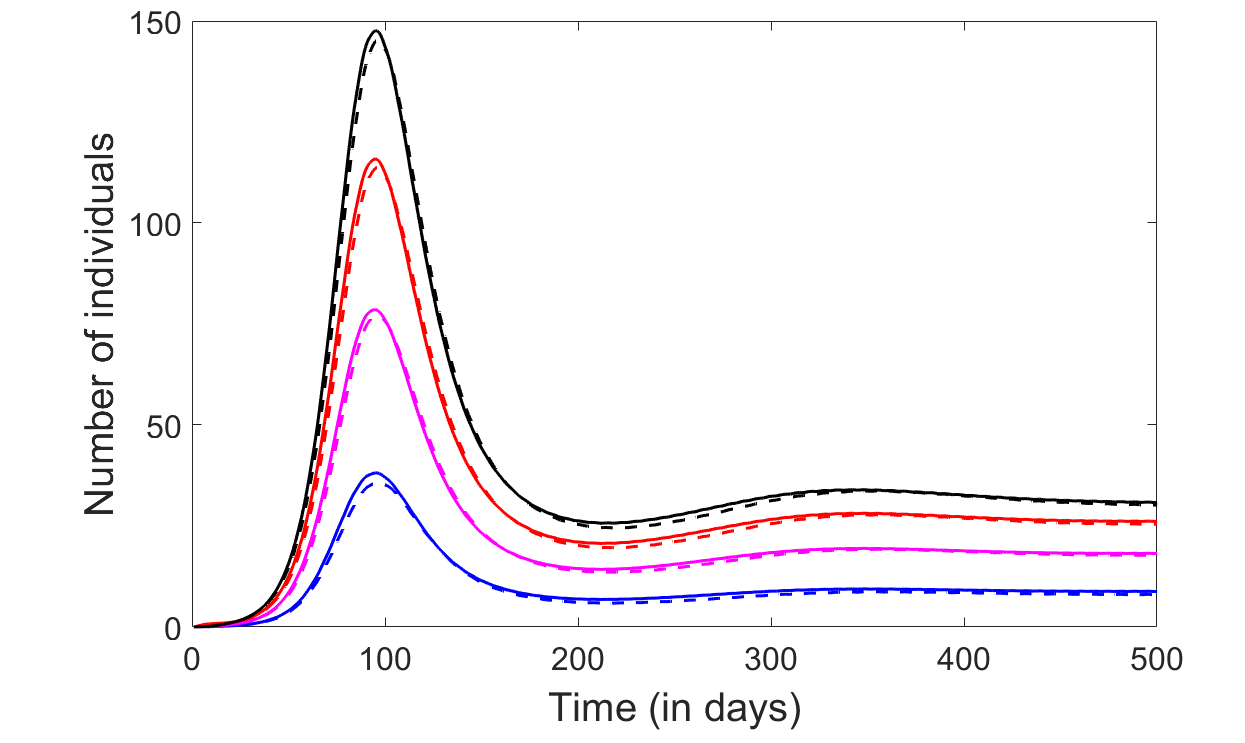}
\includegraphics[width=0.32\textwidth]{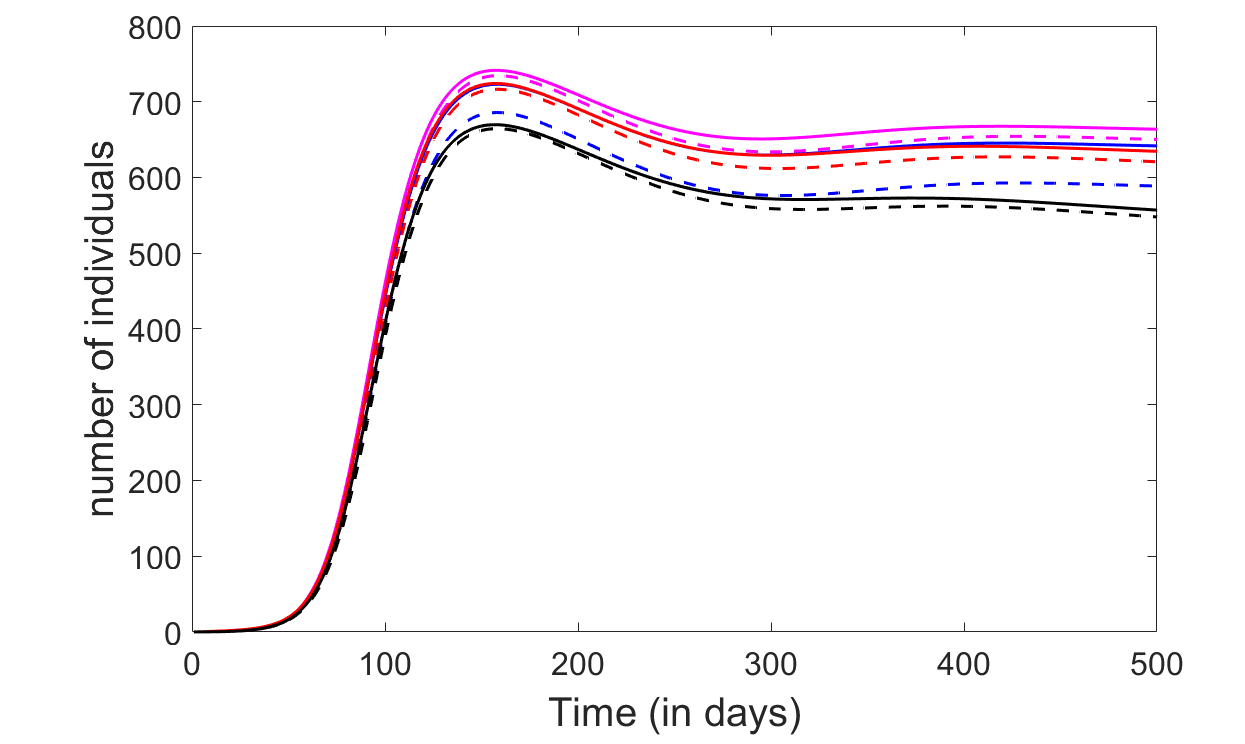}
\includegraphics[width=0.32\textwidth]{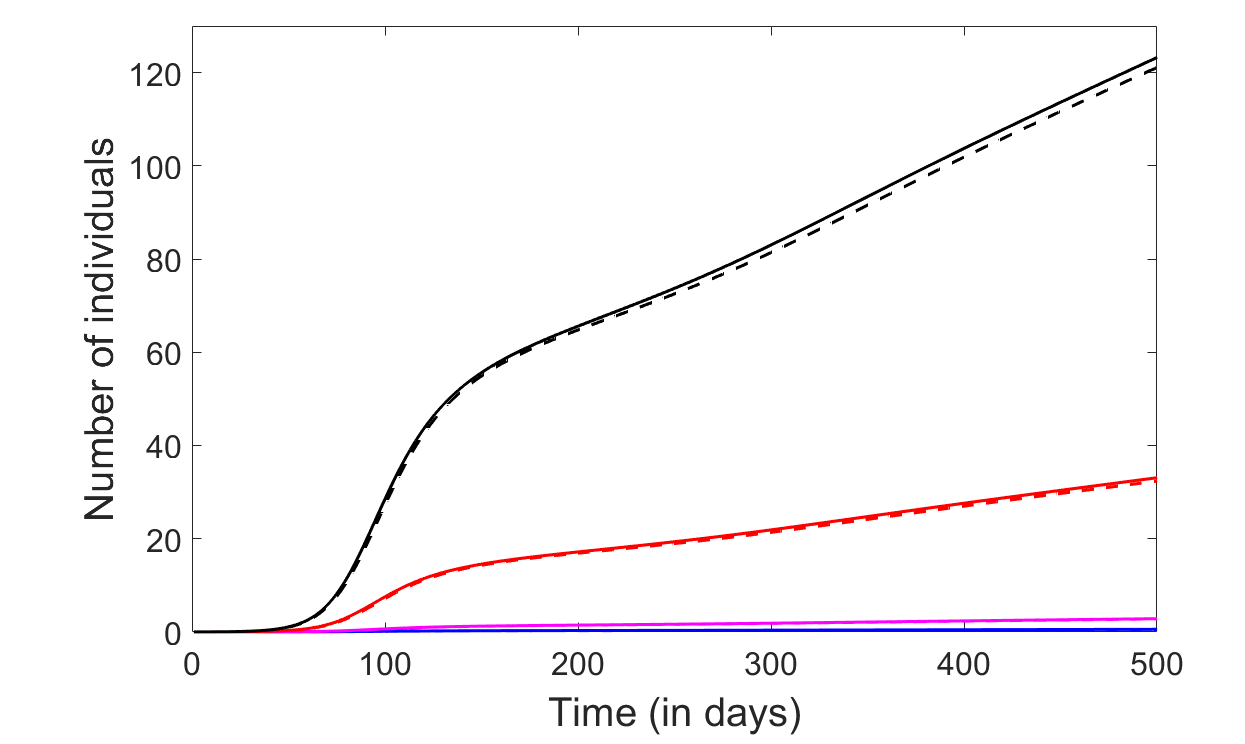}
\end{center}
\caption{\small {\it {\bf Effect of the school district closure on the systems dynamics.} The left panel represents the infected compartment, the center panel shows the recovered compartment and the right panel, the fatalities. Each age group is represented in one color: Children (blue), Young adults (pink), Adults (red) and Elderly (black). The solid curves illustrate the solutions in absence of closures; the dashed curves show the effect of the school closure imposed 15 days after infection.  }}
\label{schools_close}
\end{figure}

\begin{figure}[h!]
\begin{center}
\includegraphics[width=0.32\textwidth]{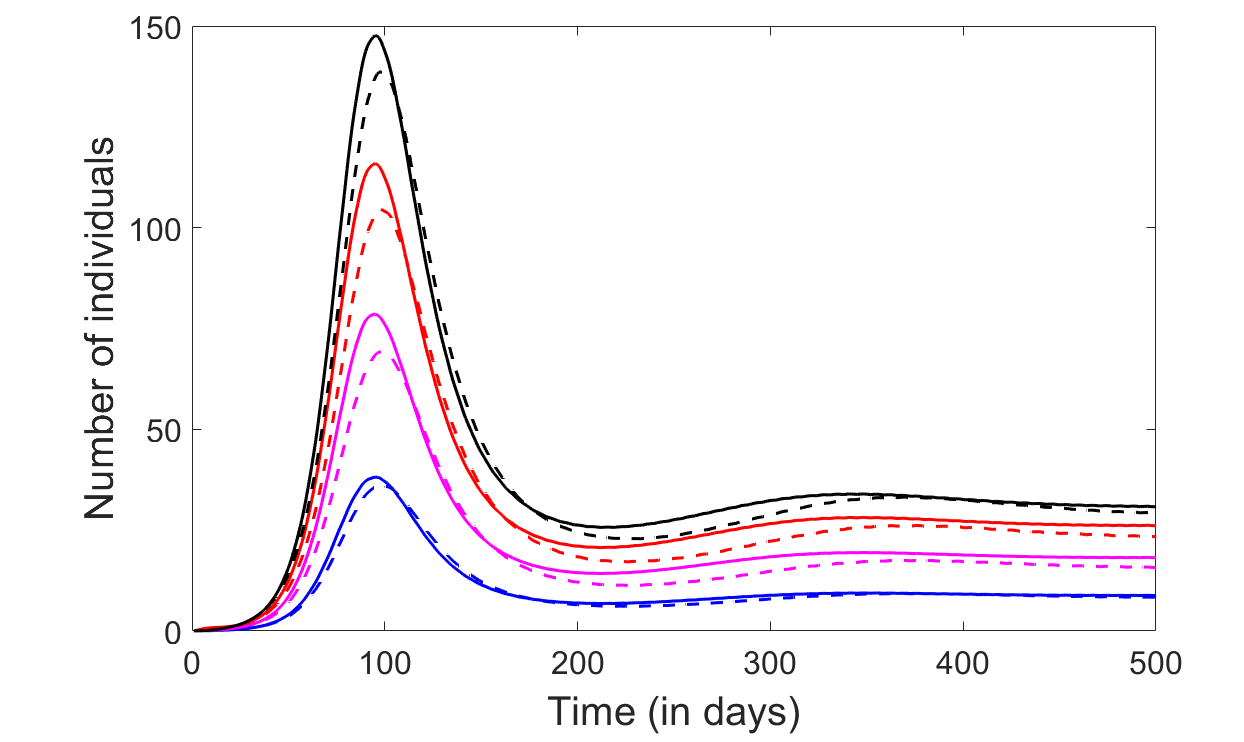}
\includegraphics[width=0.32\textwidth]{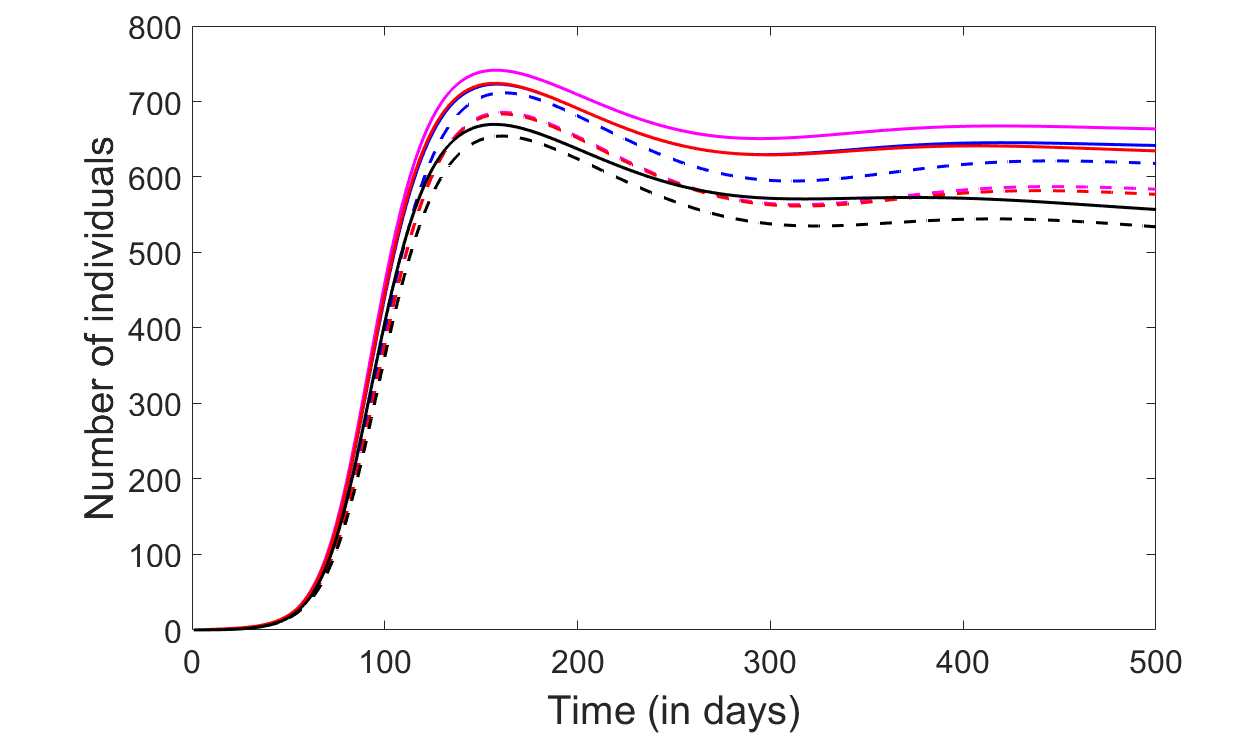}
\includegraphics[width=0.32\textwidth]{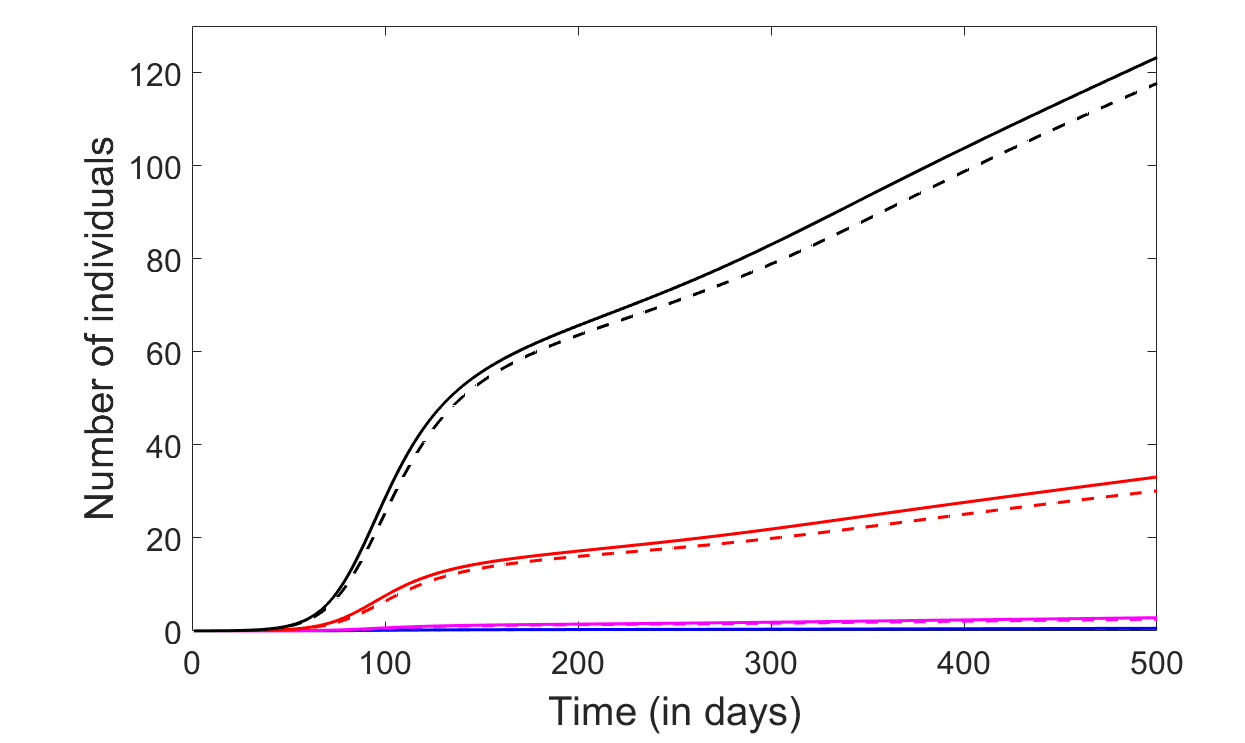}
\end{center}
\caption{\small {\it {\bf Effect of bar closures on the system dynamics.} The left panel represents the infected compartment, the center panel shows the recovered compartment and the right panel, the fatalities. Each age group is represented in one color: Children (blue), Young adults (pink), Adults (red) and Elderly (black). The solid curves illustrate the solutions in absence of closures; the dashed curves show the effect of the bars closure implemented 25 days after infection. }}
\label{bars_close}
\end{figure}

The first response strategies applied in response to the COVID 19 epidemic were focused primarily on mandated closures (i.e., mobility restrictions) rather than on social distance requirements (e.g., the recommendations on wearing masks were rather mixed over the first few weeks of the outbreak). Around the US, university campuses were among the first to dramatically limit their traffic (to virtually negligible), followed by closing school districts and, within the following two weeks, by additionally shutting down activity in restaurants, bars and entertainment venues. In Figures~\ref{campus_closes},~\ref{schools_close} and~\ref{bars_close}, we illustrate the effect of independently closing these three locations to all traffic. The closures are introduced in the model at the approximate time in the course of epidemic development as they were implemented in most New York State communities; the timeline of the response varied widely between states.

The model predicts that, even when implemented early in the process, the effect of each of the three measures taken separately is quite limited in size, and localized to the age groups directly affected. Closing the campus produced a small, but noticeable effect in Young adults, lowering (but not visibly shifting) the infection curve, as well as reducing the recovered compartment, and slightly diminishing fatalities primarily within the specific age group of Young adults. The effects on other age groups appeared negligible, hence this cannot be viewed as an efficient control measure to apply in and of itself. This analysis is restricted to the social dynamics within the community, limiting the campus interaction between Young adults and the other age compartments (only indirectly affected by the campus closure). The simulation did not include Young adult exodus from the community upon campus closure, or increased traffic to other locations to compensate, which will both be addressed in our future work on coupled communities.

Similarly, closing schools produced a very limited observed effect in our model, and primarily affected the Children age group (lowering infections, recoveries, and the already very low number of fatalities). However, this effect may be significantly underestimated by the nature of our model, which limits the interaction of children with other age compartments, especially that which occurs in a typical household. While young adults in a community of primarily college students are likely to observe more separation once in their own homes (single living), children will interact with their families, even when completely separated from others in the community. COVID spread from children to adults, and within the same household are currently subject to intense scientific debates. However, these dynamics cannot be intrinsically captured by a simple, single community SEIR model, and would require further compartmentalization, or agent-based modeling, in future work.

Closing the bars has a qualitatively different effect, due to both the specific social dynamics and also to the nature of the interactions being restricted (a place with high exposure parameter $\beta$ being eliminated). Figure~\ref{bars_close} suggests that even a slightly delayed closure (25 days) visibly lowers the infection curve in all age groups (including Children, who do not directly attend bars in our model), slightly shifts the infection curves to the right, diminishes the asymptotic number of fatalities and lowers the number of individuals in recovery, for all age groups.

\begin{figure}[h!]
\begin{center}
\includegraphics[width=0.32\textwidth]{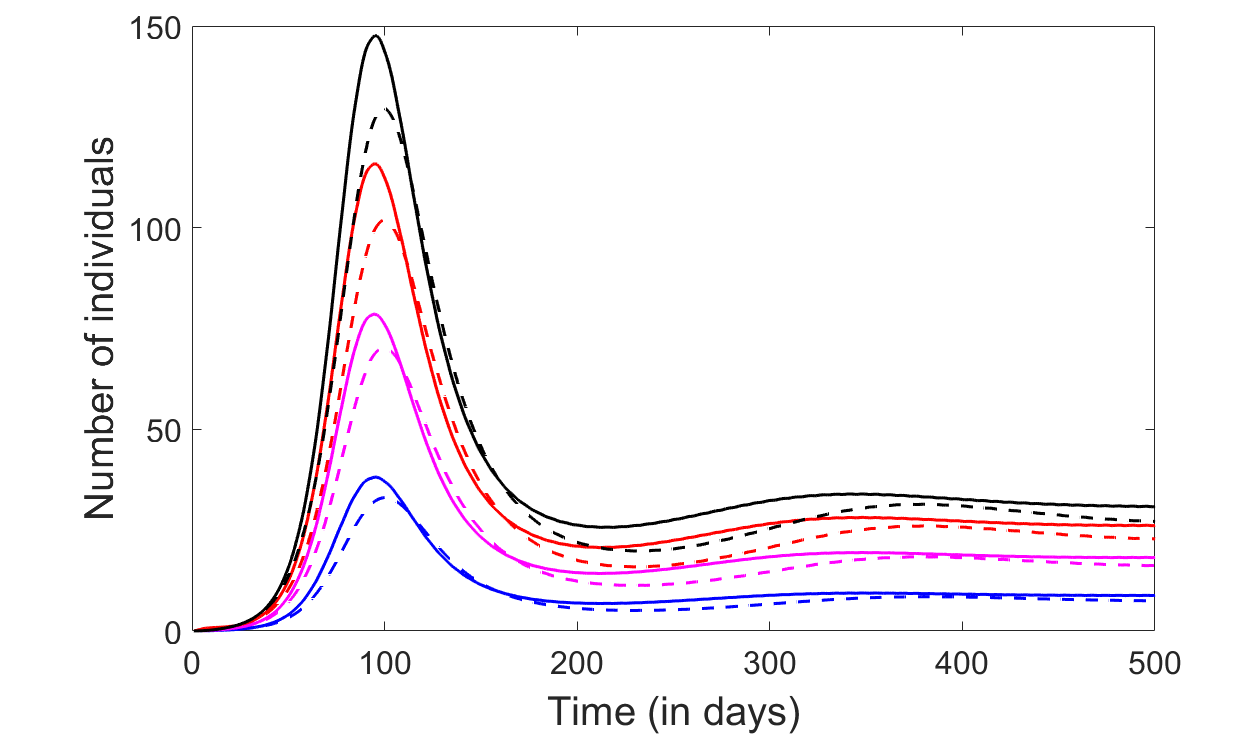}
\includegraphics[width=0.32\textwidth]{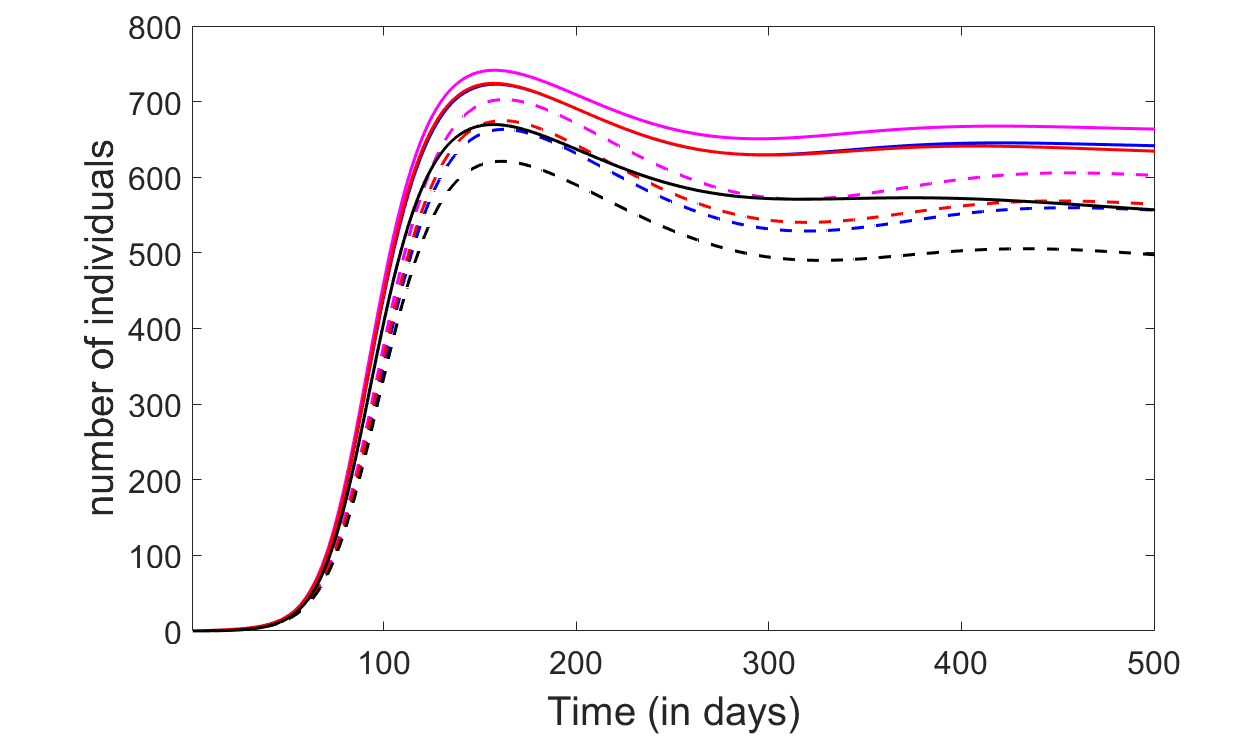}
\includegraphics[width=0.32\textwidth]{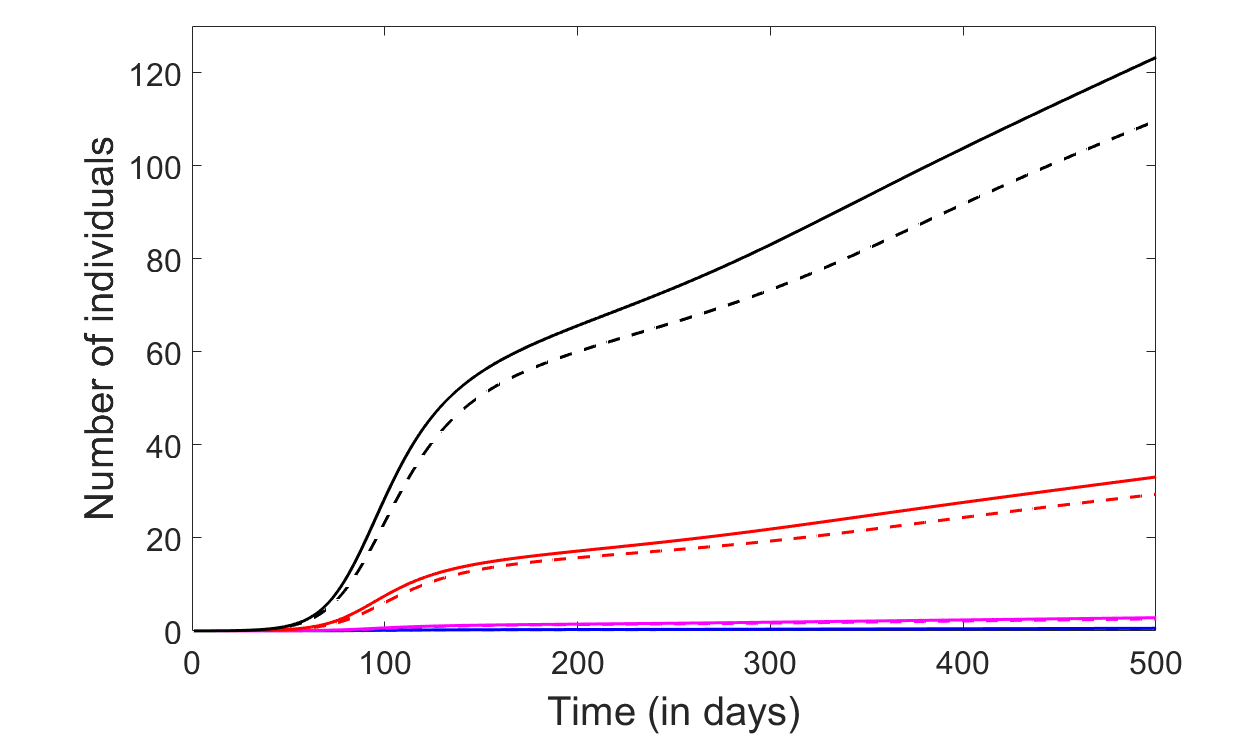}
\end{center}
\caption{\small {\it {\bf Effect of shutting down attendance to religious services.} The left panel represents the infected compartment, the center panel shows the recovered compartment and the right panel, the fatalities. Each age group is represented in one color: Children (blue), Young adults (pink), Adults (red) and Elderly (black). The solid curves illustrate the solutions in absence of closures; the dashed curves show the effect of the church gathering restrictions implemented 25 days after infection. }}
\label{churches_close}
\end{figure}

Link tracing the infection networks in various geographic regions suggests that significant community contamination has been occurring at religious / spiritual gatherings (church and synagogue services, weddings, etc.) A notorious patient \#31 in Korea reportedly infected in February over 1,160 individuals in her Shincheonji Church of Jesus congregation by attending service twice after onset of symptoms~\cite{Korea_cluster}. One of the initial large spikes in infections was observed in New York City the week of March 17, due to a quick spreading in tightly knit Jewish communities in Westchester~\cite{Westchester_cluster} and Brooklyn~\cite{Brooklyn_cluster}. Many churches have been heavily criticized for continuing to observe specific rites which contravened the hygiene and distancing directives given in conjunction with the pandemic (people in close proximity, using the same spoon and cup for communion~\cite{church_spoon,church_cup}). Starting with March 23rd, many New York State churches announced suspension of services, or no-attendance services~\cite{church_closures}.

Our model builds in a relatively small daily church attendance, but also one day per week focused specifically on such community activity, reflected into doubled exposure rates at the respective location (from $2\beta$ to $4\beta$) due to proximity to others, sharing utensils, etc. We experimented by first enforcing stricter separation specifically during service (lowering the exposure parameter only for the weekly gatherings), which was inefficient. We then completely shut down attendance (by updating the mobility array) at day 25 from the original infection. Figure~\ref{churches_close} shows that completely closing down these events produces a qualitatively similar, but larger effect to that of the bar and restaurant shutdown, with a lowering and slight shift in infection curves; it was also accompanied by decreasing fatalities across the board for all ages.

\begin{figure}[h!]
\begin{center}
\includegraphics[width=0.32\textwidth]{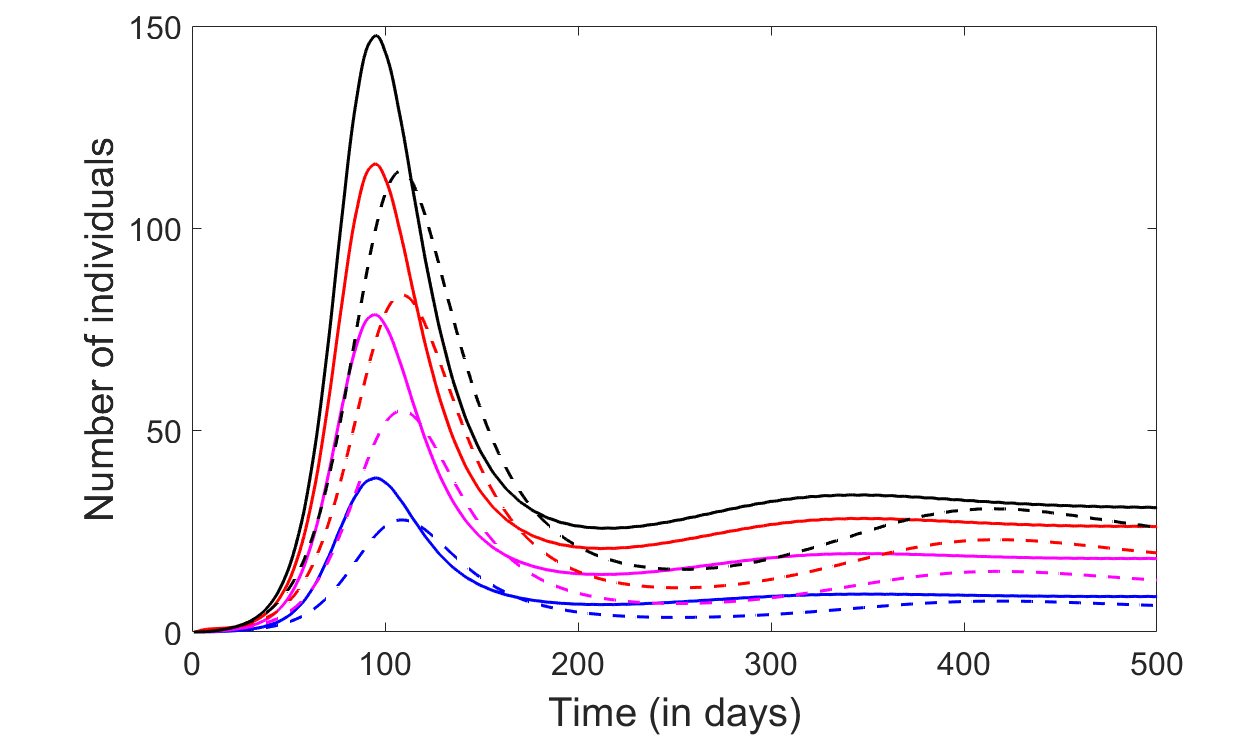}
\includegraphics[width=0.32\textwidth]{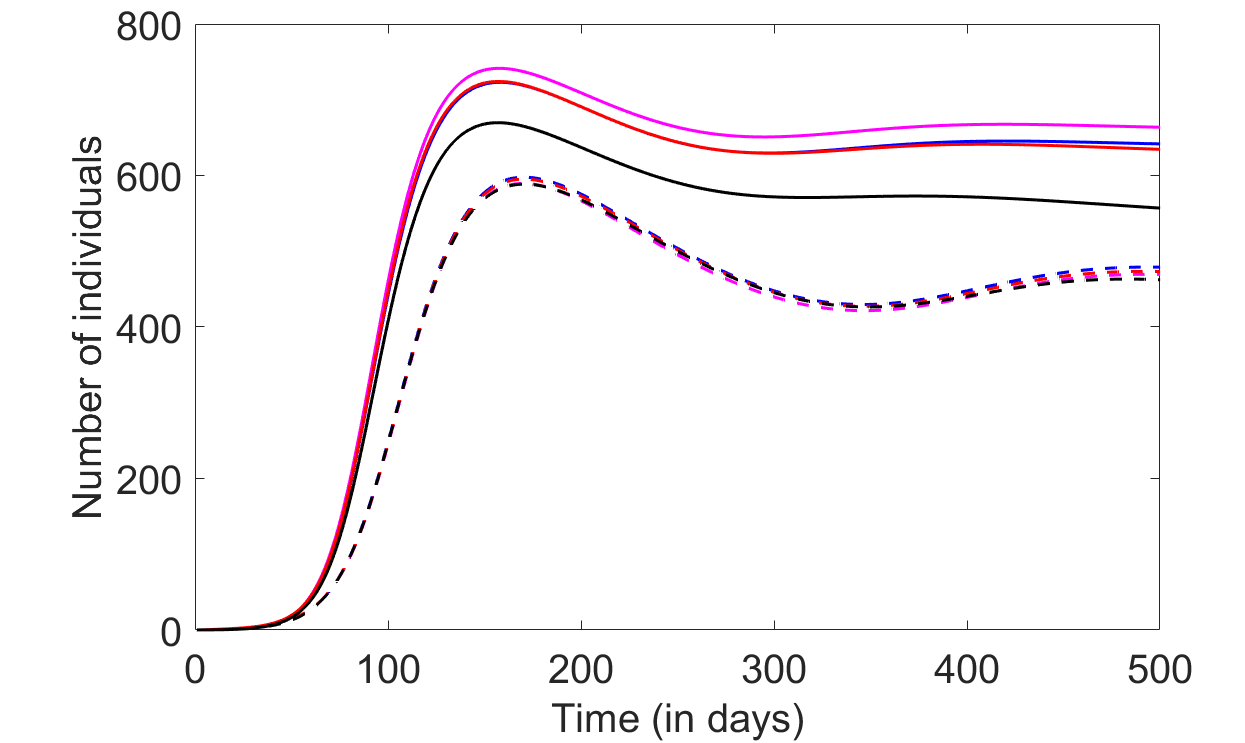}
\includegraphics[width=0.32\textwidth]{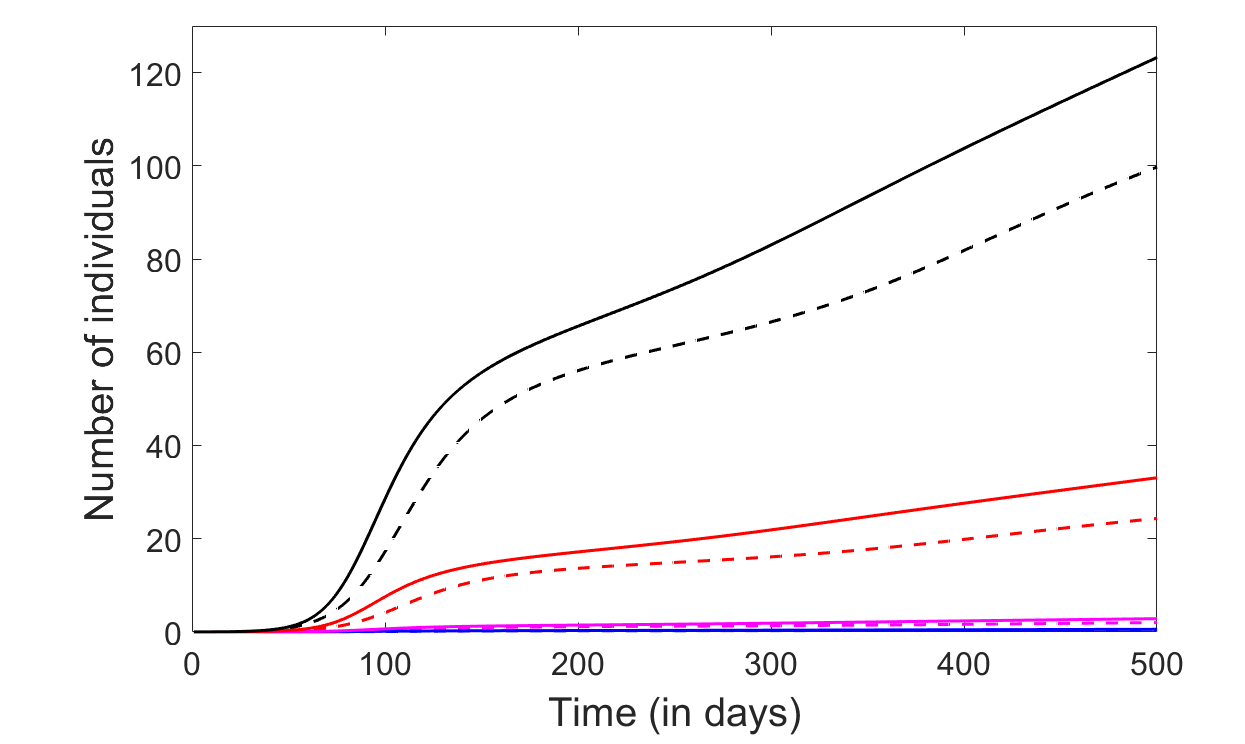}
\end{center}
\caption{\small {\it {\bf Cumulative effect of realistic closing procedures on the system dynamics.} The left panel represents the infected compartment,  the center panel shows the recovered compartment and the right panel, the fatalities. Each age group is represented in one color: Children (blue), Young adults (pink), Adults (red) and Elderly (black). The solid curves illustrate the solutions in absence of closures; the dashed curves represent the cumulated effects of closing the campus at day 10, the school district at day 15, the bars and churches at day 25 (the approximate timeline of the real life implementations of these closures in New York State).}}
\label{combined_close}
\end{figure}

While each of these closures considered independently produced a relatively small effect on curbing infection (hence the number of people in potential need of treatment), as well as on shifting the epidemic timeline -- they become more efficient when applied in combination. Figure~\ref{combined_close} illustrates their compounded contribution to ``flattening the curve,'' and to a significant reduction of infection, number of recovered individuals and fatalities in all age groups. In the context of this outbreak potentially challenging our health care capacity, the decrease in the infection maximum value is crucial, but the delay can also be of great practical importance, allowing time for a more prepared medical response to the occurrence of this maximum. The significant reduction of the Recovered compartment is also clinically relevant, in light of COVID infection being associated with increased risk for potentially serious and long-term health problems (the evidence of which is gradually starting to emerge, even in completely asymptomatic individuals).

\begin{figure}[h!]
\begin{center}
\includegraphics[width=0.32\textwidth]{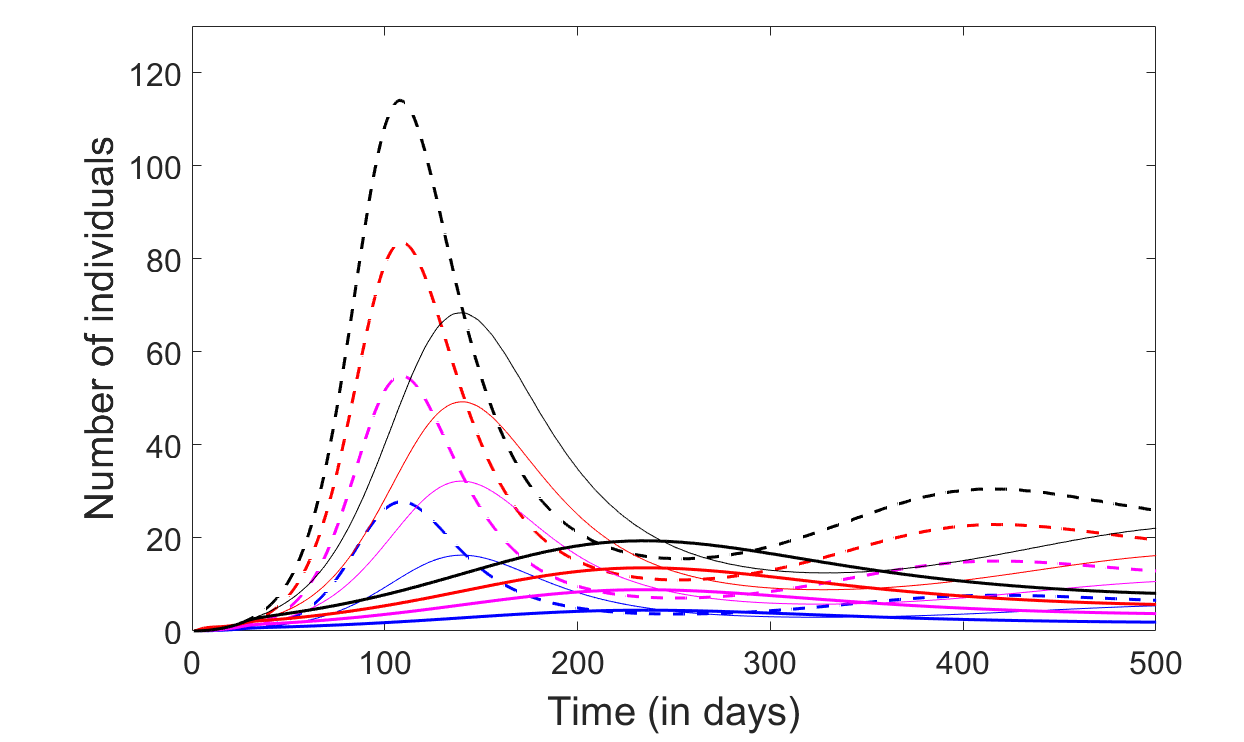}
\includegraphics[width=0.32\textwidth]{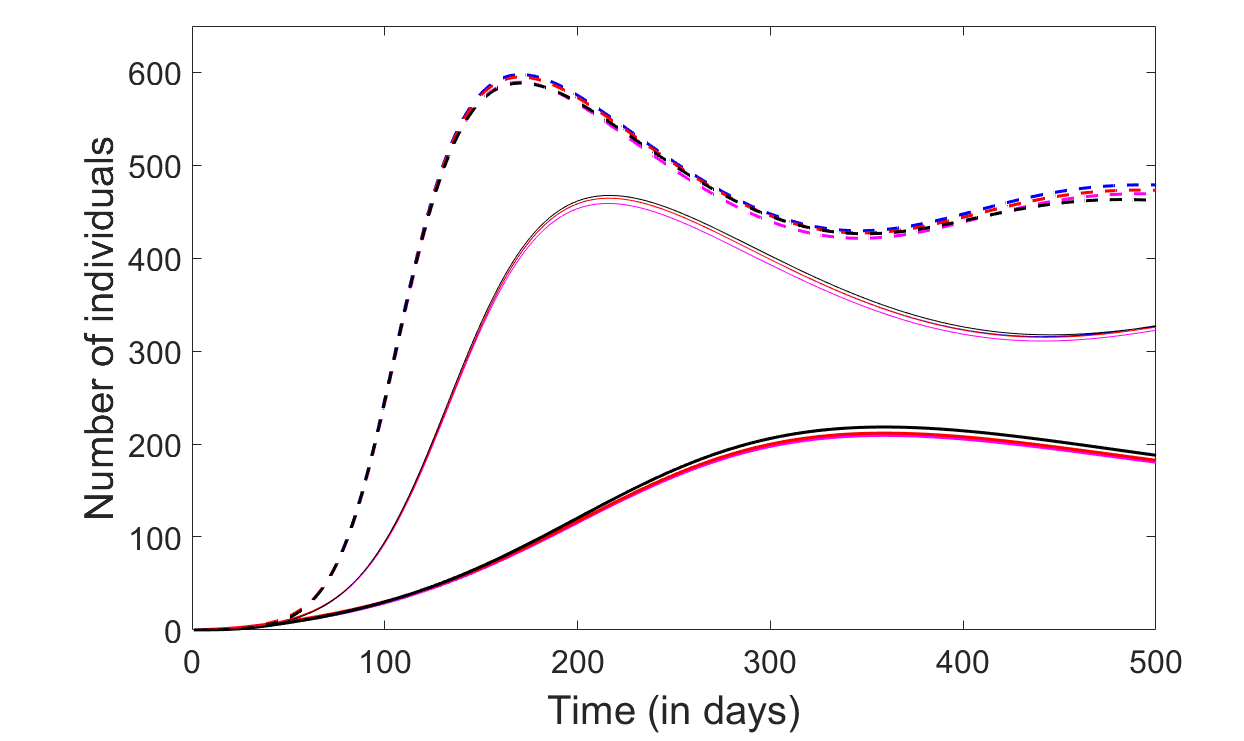}
\includegraphics[width=0.32\textwidth]{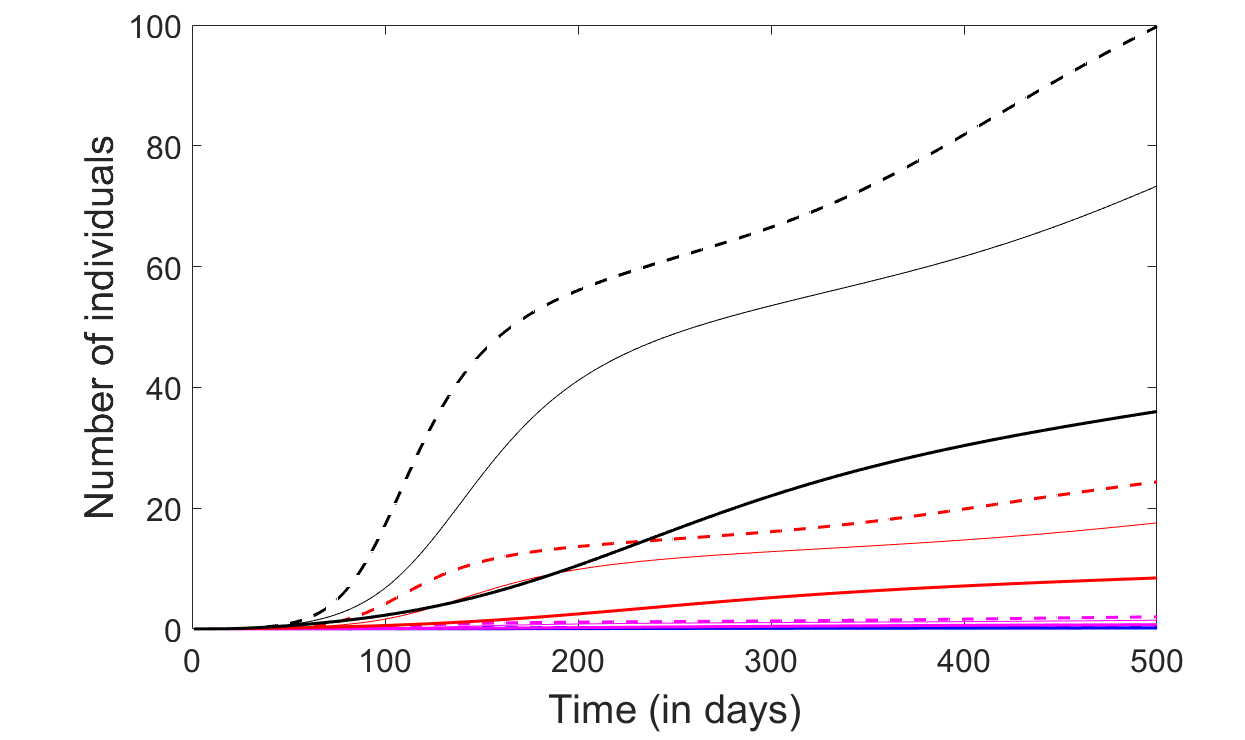}
\end{center}
\caption{\small {\it {\bf Effect of exercising social distancing in addition to the existing shutdowns.} The left panel represents the infected compartment, the center panel the recovered, and the right panel, the fatalities. Each age group is represented in one color: Children (blue), Young adults (pink), Adults (red) and Elderly (black). The dashed curves illustrate the original predictions; the thin solid curves represent the evolution of the system when the value of the exposure parameters were decreased by 20\% of the original values, to reflect the effect of social distancing at all destinations; the thick solid curves represent a deeper, 40\% reduction of $\beta$ values.}}
\label{additional_distance}
\end{figure}

Overall, our model predicted that, in combination, closing access to specific destinations can act as a first step, but could not in and of itself suppress the epidemic outbreak (e.g., the fatality curve is still increasing at the end of the 500 day observation interval, as infection persists and even launches into smaller subsequent waves). That is because, even with completely shutting down mobility to specific venues, exposure will continue to occur at destinations associated with essential needs (food, medicine, medical care), as well as in the residential space. This suggests that control of the outbreak can only be obtained by following an efficient hygiene and social distancing protocol when interacting in such spaces. Using our model, we explored the effects of practicing social distancing {\it in all locations}, including in the residential space, in addition to the closure and isolation measures that had already been implemented. We modeled this effect by reducing the exposure factor $\beta$ in all locations, as shown in Figure~\ref{additional_distance}. The simulation suggests that social distancing represents a necessary and efficient strategy in curbing the effects of the outbreak. Practically speaking: directly lowering exposure at the remaining locations (by allowing ample personal space, wearing masks, and by showing caution and good hygiene when interacting with various exposed surfaces) is predicted to be very efficient, especially in conjunction with the existing reduced mobility of the individuals to specific locations. While any degree of social distancing leads to improvements in the outcome, notice that a 20\% reduction in $\beta$ has a weaker effect, and does not completely eliminate the potential for secondary waves; in contrast, a 40\% reduction is a very efficient control measure.

\begin{figure}[h!]
\begin{center}
\includegraphics[width=0.32\textwidth]{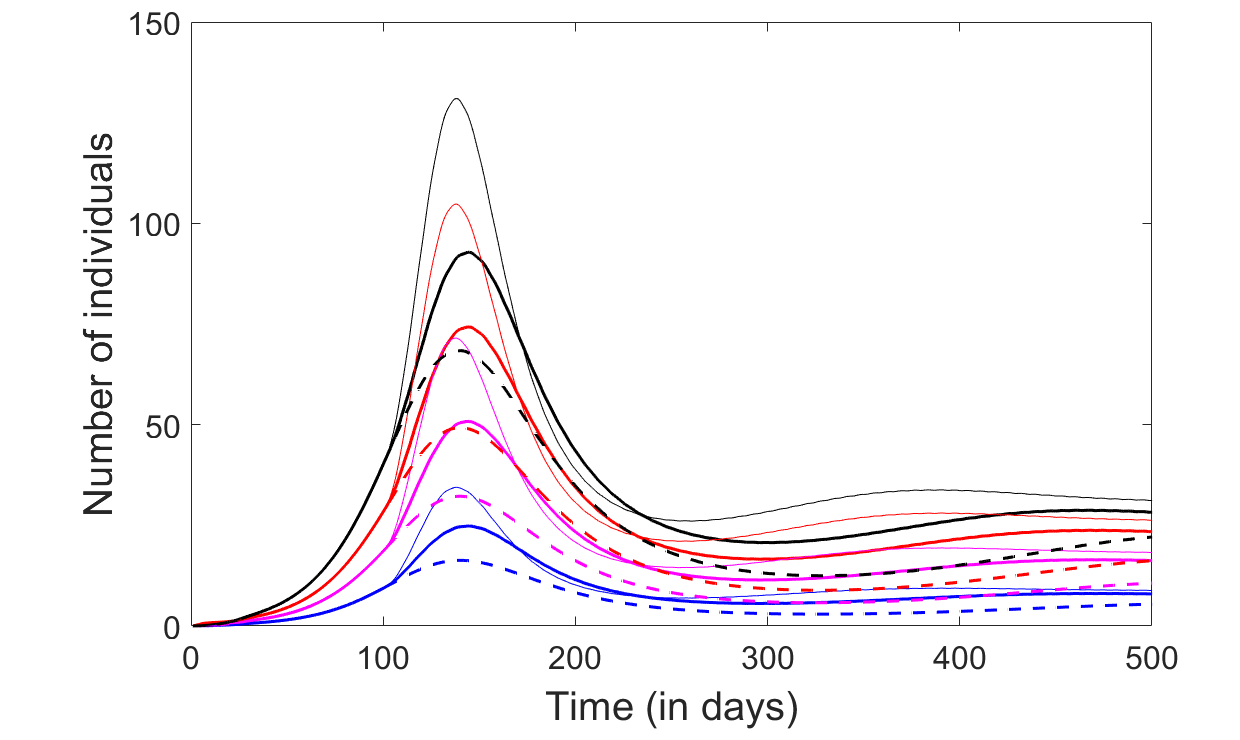}
\includegraphics[width=0.32\textwidth]{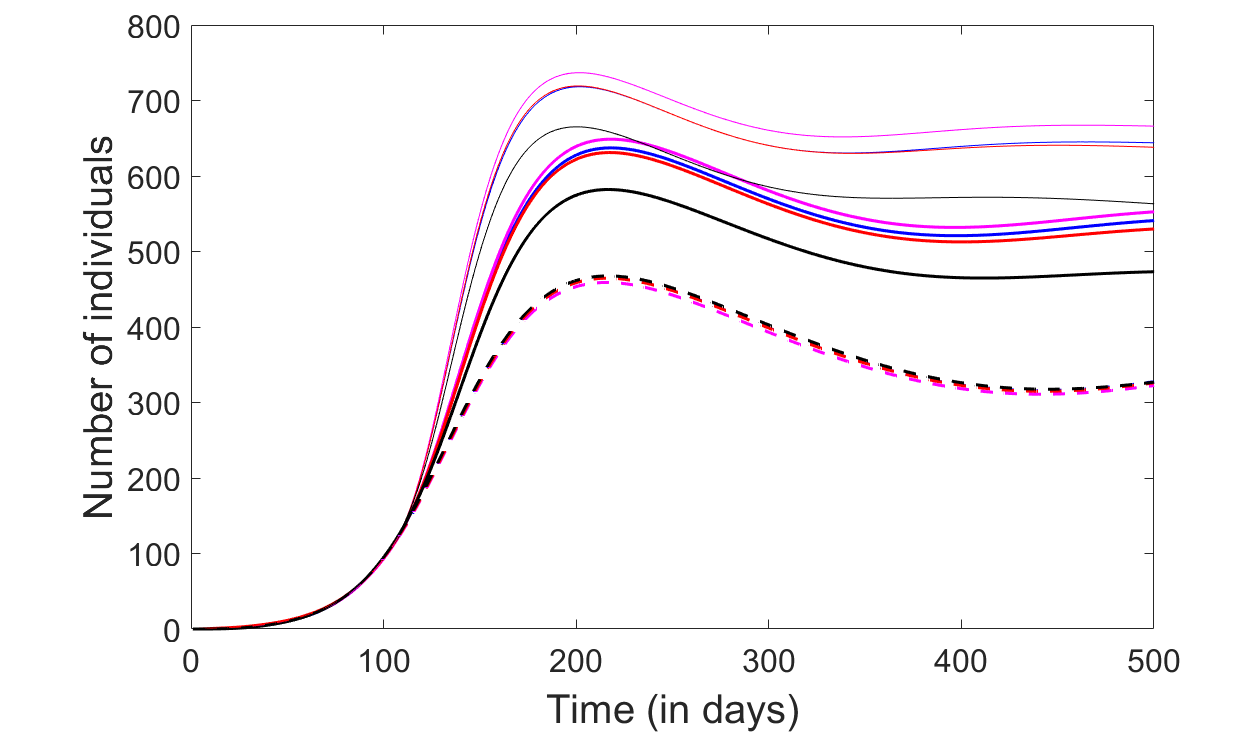}
\includegraphics[width=0.32\textwidth]{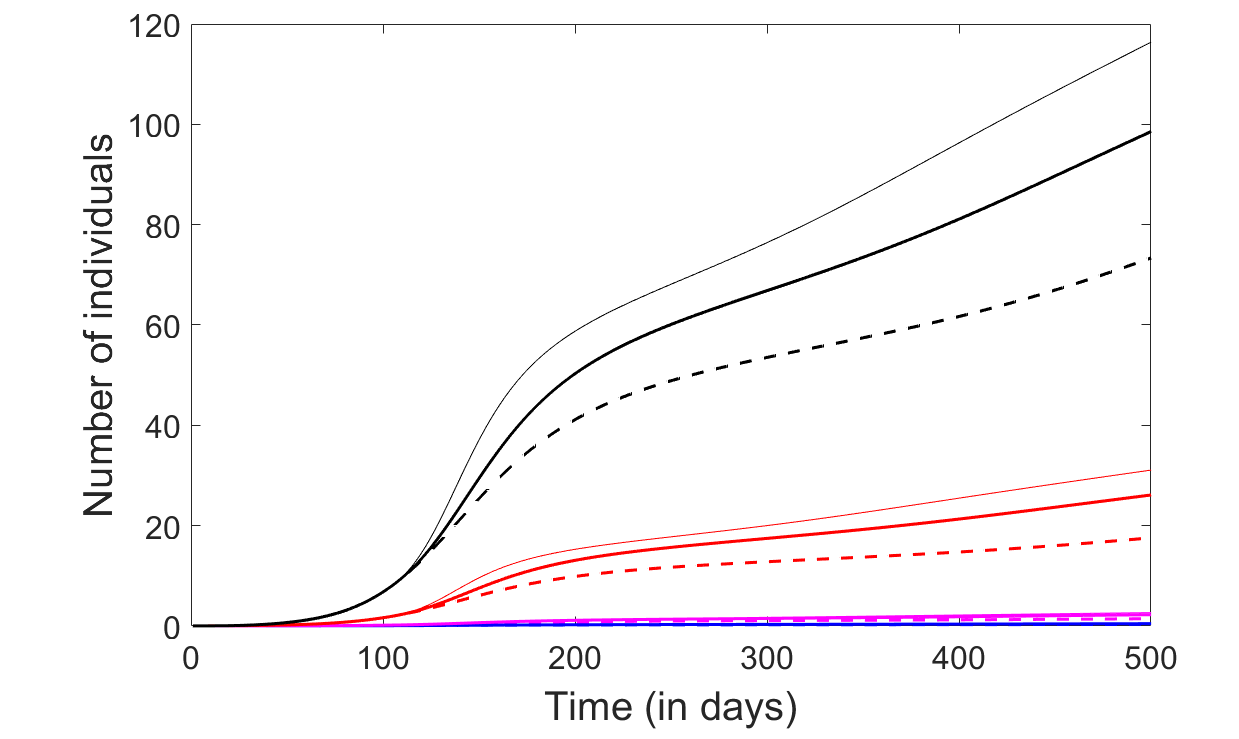}\\
\includegraphics[width=0.32\textwidth]{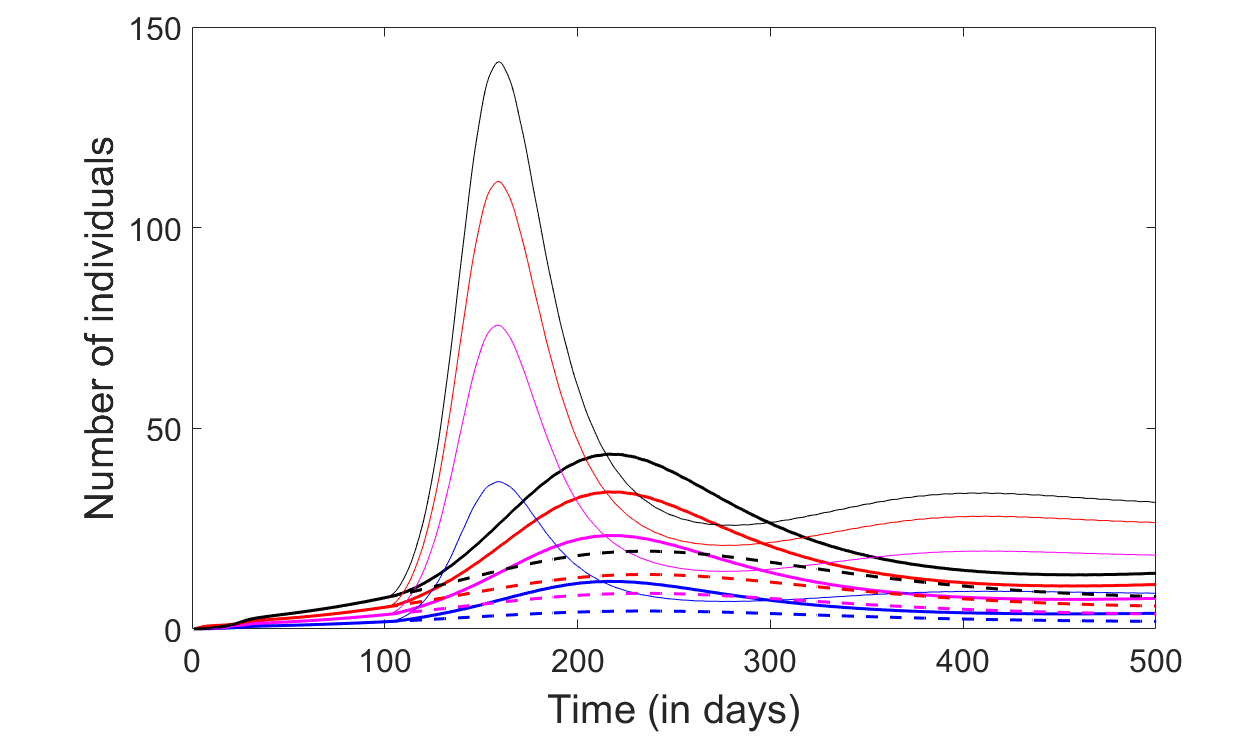}
\includegraphics[width=0.32\textwidth]{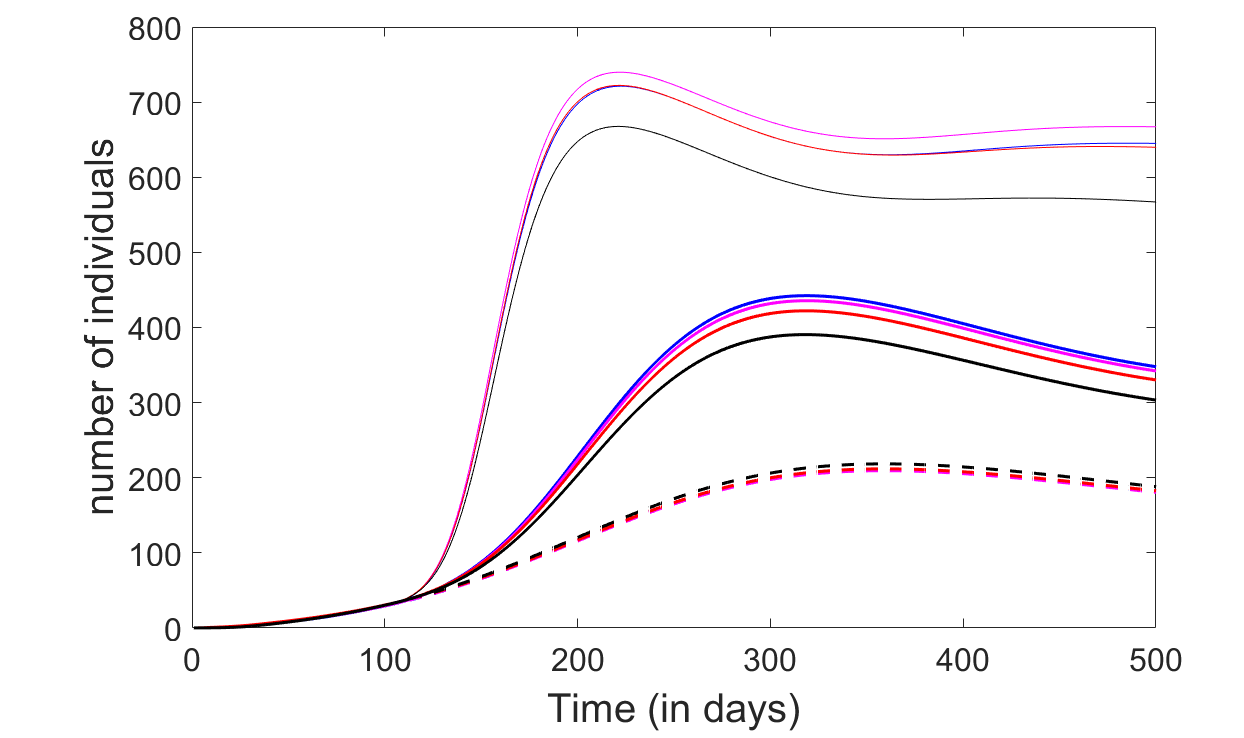}
\includegraphics[width=0.32\textwidth]{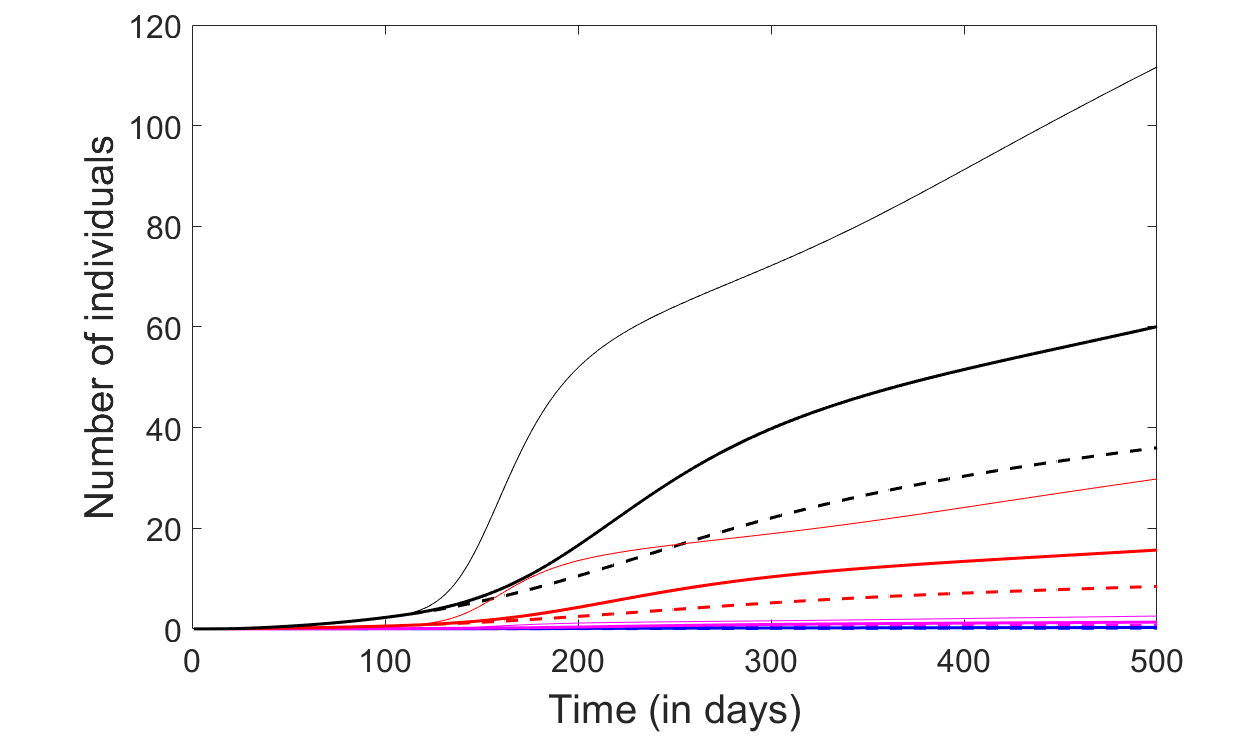}\\
\end{center}
\caption{\small {\it {\bf Effect of lifting closures and relaxing social distancing after 100 days.} From left to right, the panels represent the infected, the recovered and the fatalities compartments. Each age group is represented in one color: Children (blue), Young adults (pink), Adults (red) and Elderly (black). The dashed curves illustrate the prediction with closures and social distancing in place. The thin solid curves illustrate the prediction with both closures and social distance restrictions lifted. The thick solid curves represent a scenario in which closures are lifted, but social distancing is maintained. In the top panels, $\beta$ is reduced by 20\% with social distancing; in the bottom panels, it is reduced by 40\%.}}
\label{relax1}
\end{figure}

Our final analysis consisted of using the model to implement different reopening strategies and timelines. In Figure~\ref{relax1}, we compare the outcomes of three different scenarios with the same timeline (reopening at 100 days from the start of the epidemic, corresponding broadly to the point when a few states started the reopening process in the US). In the top panels, we implemented weaker social distancing (translated in the model by a \% reduction in the exposure factor $\beta$ at all accessible locations); the bottom panels represent the outcome of the same reopening schedule, but with a stricter, 40\%, social distancing.

One extreme scenario (dotted curves), is to postpone reopening indefinitely, and continue imposing the mobility restrictions, as well as strict (e.g., 40\%) social distancing measures. This scenario delivered, of course, the best outcome, but the conditions are unrealistic, and undesirable in the context of optimizing epidemiological safety versus economic survival. The other extreme (dashed curves) was to simultaneously withdraw both mobility and social distancing restrictions, and fully reopen on day 100. This was, of course, the riskiest plan, which lead to only slightly flattening the first epidemic wave (now delayed by approximately 100 days). As a middle ground, we considered the scenario of restoring mobility on day 100, but preserving social distancing.

Even with the weaker social distancing (top panels), the model predicted clear quantitative advantages to taking the middle ground versus the riskier path: the infections were substantially curbed (even though not as much as with perpetual closures), the recovered population and fatalities were decreased in all age groups. When considering the stronger social distancing scheme (bottom panels), the outcomes of the three strategies were qualitatively different. With the riskier scenario, the infection wave was again delayed, and in fact even less efficiently suppressed. In contrast, the extreme scenario of perpetual closures and social distancing efficiently controlled the infection, both in terms of size, and of possibility of future waves (subsequent peaks). Interestingly, however, a similar effect was obtained by merely sustaining the strict social distancing, but otherwise restoring all social mobility to destinations. While restoring mobility slightly raised all curves, it did not do so significantly, and, more importantly, it maintained the control of the epidemic (convergence without further damped oscillations).

\begin{figure}[h!]
\begin{center}
\includegraphics[width=0.32\textwidth]{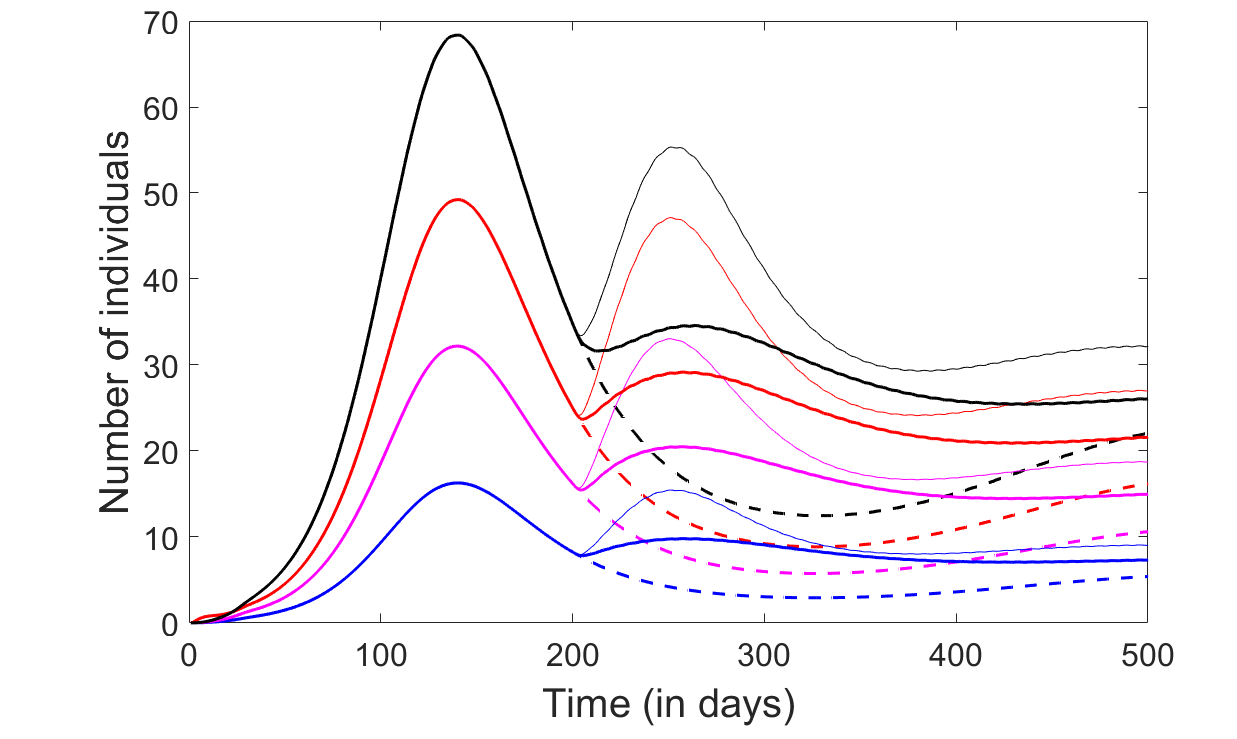}
\includegraphics[width=0.32\textwidth]{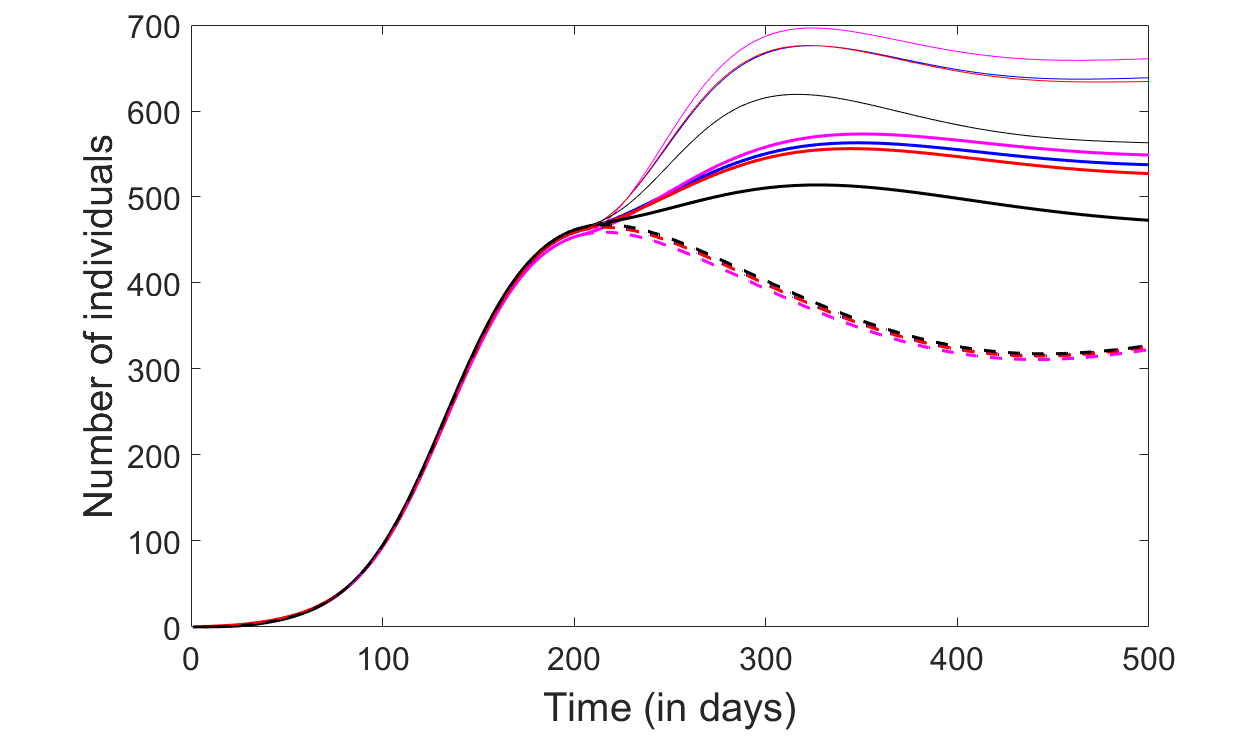}
\includegraphics[width=0.32\textwidth]{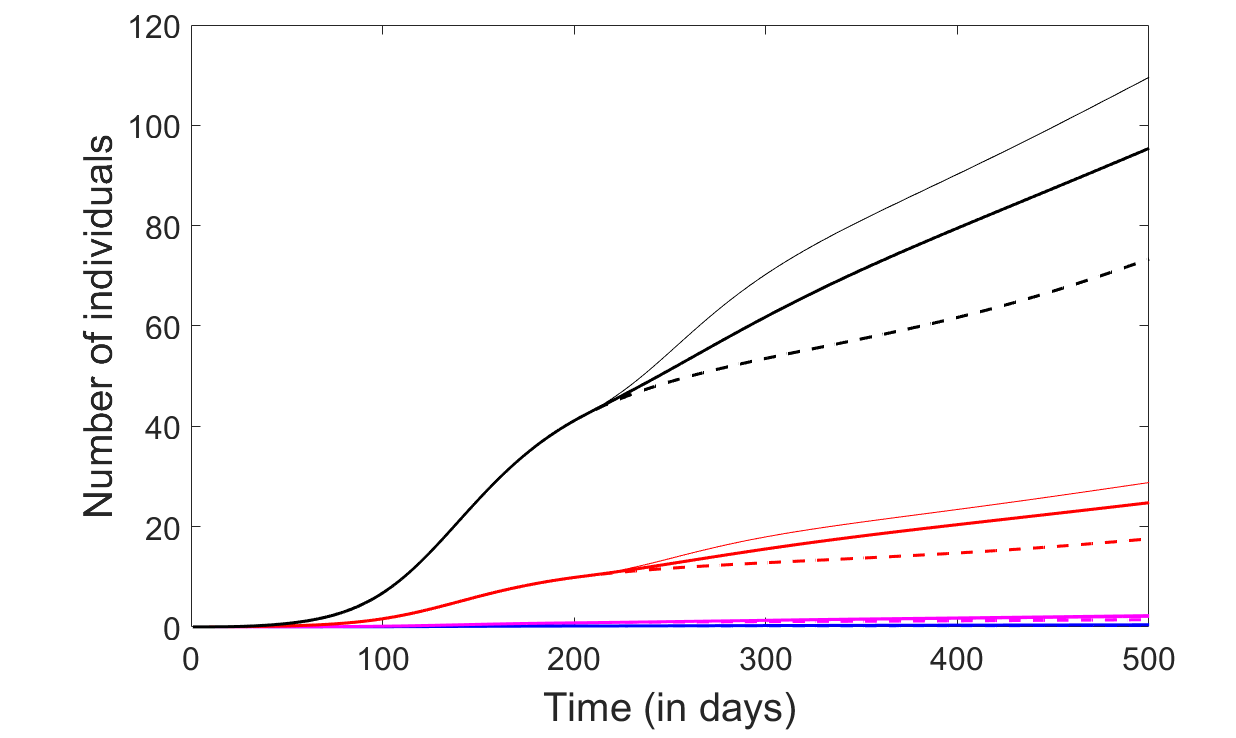}\\
\includegraphics[width=0.32\textwidth]{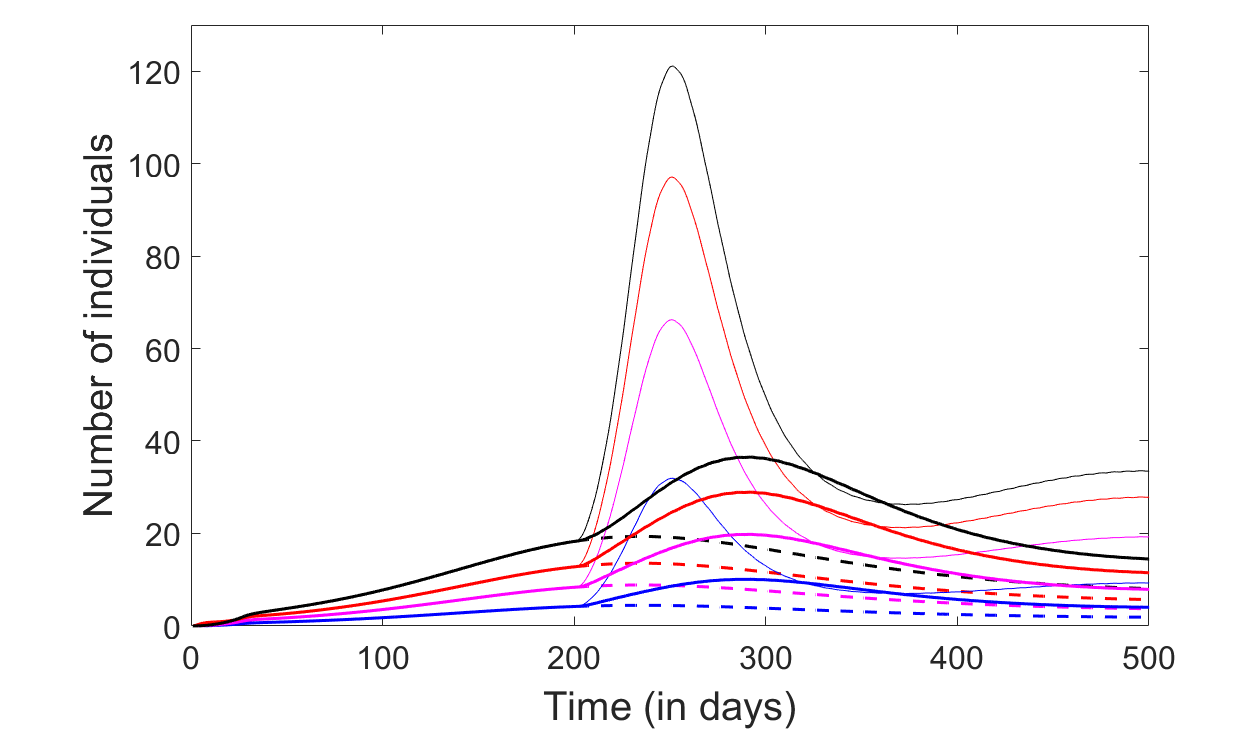}
\includegraphics[width=0.32\textwidth]{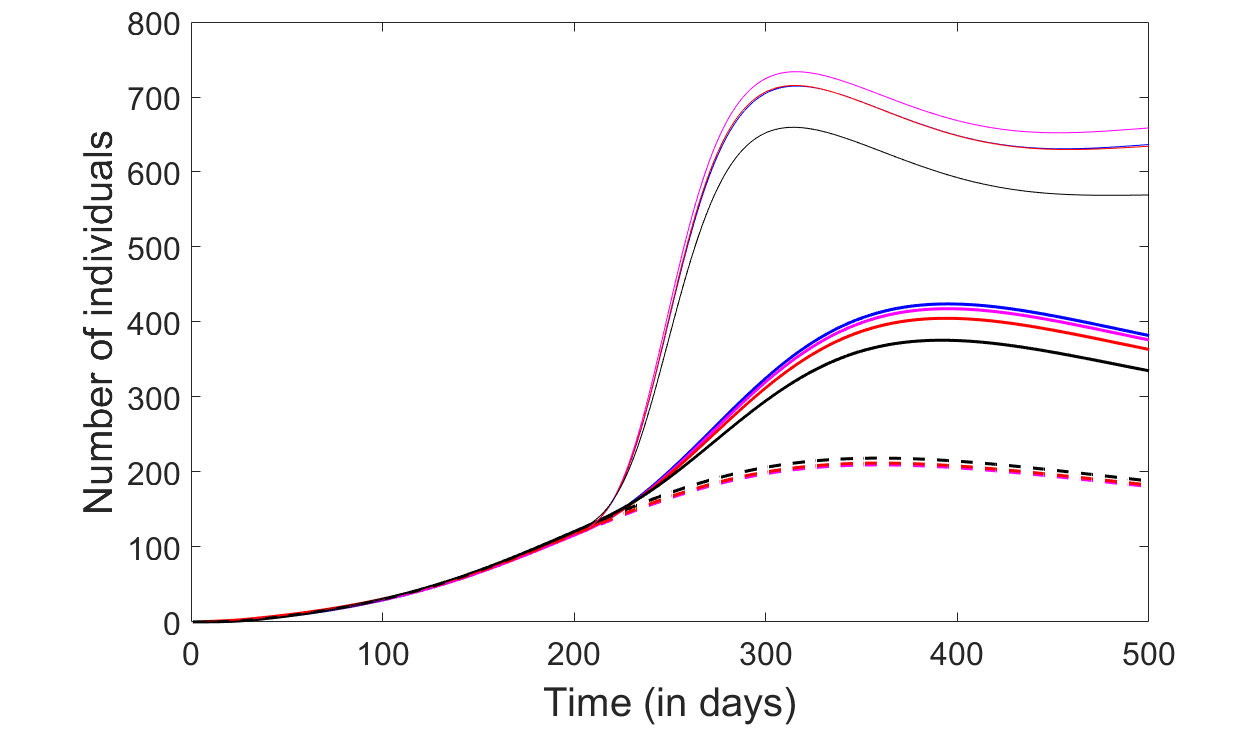}
\includegraphics[width=0.32\textwidth]{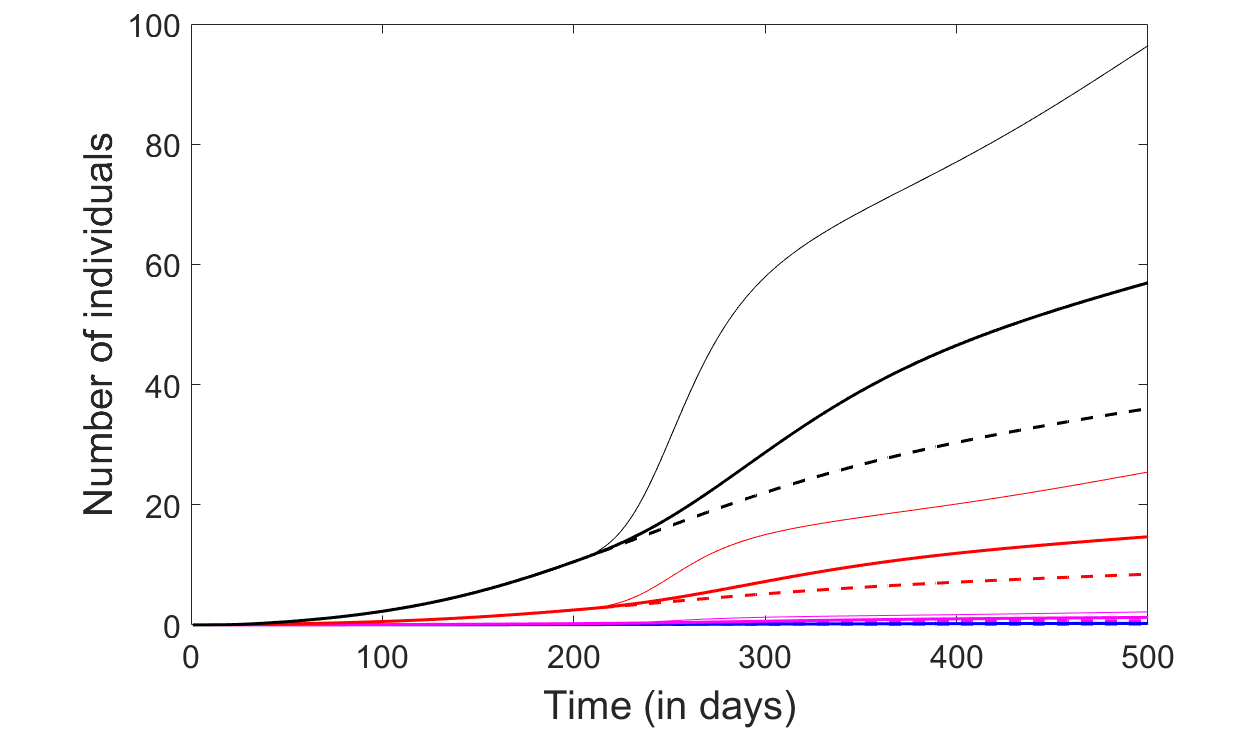}\\
\end{center}
\caption{\small {\it {\bf Effect of lofting closures and relaxing social distancing after 200 days.} From left to right, the panels represent the infected, the recovered and the fatalities compartments. Each age group is represented in one color: Children (blue), Young adults (pink), Adults (red) and Elderly (black). The dashed curves illustrate the prediction with closures and social distancing in place. The thin solid curves illustrate the prediction with both closures and social distance restrictions lifted. The thick solid curves represent a scenario in which closures are lifted, but social distancing is maintained. In the top panels, $\beta$ is reduced by 20\% with social distancing; in the bottom panels, it is reduced by 40\%.}}
\label{relax2}
\end{figure}

In Figure~\ref{relax2}, we compare the outcomes of the same three scenarios, but with a reopening timeline of 200 days from the start of the epidemic (corresponding broadly to the current point, for many communities in the US). Notice that, in the case of weaker social distancing (20\% reduction in $\beta$), fully reopening at a time when the infection is headed down a decreasing slope has the potential to immediately prompt a sharp second spike, which would be significantly milder if social distancing were maintained in place. As before, the combination of restricted mobility and social distancing are not efficient when applied for only the limited time (200 days), even with strict social distancing. However, prolonging social distancing by itself is enough to efficiently control the epidemic (reduce the size of infected, recovered and fatality compartments and suppress future waves), even when mobility is restored.

\section{Discussion}

\subsection{Comments on the model}

In this paper, we constructed and briefly analyzed a compartmental system of equations that captures the epidemic dynamics of the ongoing COVID 19 outbreak. In a community including people from all age groups, we first simulated numerically the effects of the social mobility restrictions that were mandated in New York State communities upon emergence of the first epidemic wave (closing of campuses, schools, restaurants, entertainment and spiritual gatherings), with timing matching approximately the average time (from the original infection in the community) when these measures were implemented in the field. We then simulated different reopening timelines and strategies, in search of an optimal combination that would restore mobility, yet maintain an adequate level of control of the epidemic.

When simulating the shutting down timeline, we analyzed the effect of each measure applied independently, and observed effects limited in size, and to specific age groups. Separately closing the campus, the schools, the bars and the churches induced limited effect on flattening the infection curve, on reducing the number of exposed and then recovered individuals, or on the overall number of fatalities. In addition, closing the campus primarily benefited young adults; closing schools impacted children; closing the bars was most effective for young adults and adults, and closing access to group religious services primarily impacted the elderly and the adults, as shown by the illustrations of infected, recovered and fatality dynamics. When simulating in the sequence and at the approximate timing at which they were implemented, the {\it combination} of all these factors significantly affected all age groups. However, while this strategy alone reduced the potency of the first infection wave, it did not efficiently curb the epidemic (our simulations show that it in fact enhanced the possibility of more pronounced secondary waves).

We further investigated the effects of exercising caution, additional hygiene and social distancing practices (such as wearing a mask, disinfecting hands, respecting physical distance to others), both at home and outside of the home, by lowering the exposure rates in all locations. Paired with social distancing, the closure strategies gained dramatically in efficiency, even when social distancing was introduced after mobility had already been reduced (which is a fair representation of the real life implementation). Stricter social distancing measures in addition to closures were associated with a significantly improved outcome.

All these simulations, however, consider the unrealistic scenario of maintaining closures in perpetuity. These were proven to be unrealistic to maintain even over a period of 3-5 months, placing a huge strain on the economy, as well as on individuals' psychological balance and livelihoods. To address the difficult question regarding the best time and conditions of reopening, one needs to weigh the advantages versus the clinical implications. We simulated different reopening strategies, while continuing to monitor in particular the same three epidemic compartments: the number of infected individuals (measuring symptomatic incidence, and the strain on the heath care system); the number of recovered individuals (measuring the virus exposures, symptomatic or not, which may lead to further COVID-related health issues in the future); the number of fatalities (measuring the overall loss of life produced by the outbreak thus far).

Our simulations considered different reopening times (from an early reopening at 100 days, during the upward progression of the first epidemic wave, to a later reopening at 200 days, after the first wave peak had time to wane). We also compared full reopening (lifting both mobility and social distancing restrictions) to the strategy of only restoring mobility, and keeping social distancing measures in place. Unsurprisingly, early full reopening augmented the first wave; later full reopening prompted a significant second wave (if the first wave had already happened), or allowed the first wave to fully develop (in case stricter conditions had kept it in check thus far). Either way, full reopening did not do justice to the efforts already invested in controlling the epidemic, and lead to effects comparable in size to those that could have been obtained with little to no control in the first place. In contrast, our model predicts that reopening with social distancing is a much more efficient strategy, which maintains a level of control of the epidemic qualitatively comparable with that obtained with no reopening. This option also builds upon the efficiency of past measures: the prognosis is better if stricter social distancing had already been practiced, and the control increases with the length of time spent in lockdown.

Curbing population mobility, locally as well as globally, is a necessary and typical first response in the face of a pandemic. However, it cannot constitute a long-term plan in and off itself, since it conflicts with economic and societal needs. Upon emergence of the COVID 19 outbreak, a lot of weight has been placed on reducing population traffic to various destinations, and a lot of controversy has occurred around the necessity of such measures. As it began to appear that the epidemic was going to stay for the long haul, there has been a change in focus around measures that can maintain control of the outbreak, yet can be more easily maintained in the long term, such as social distancing. While social distancing measures have not been free of debate, they are clearly a more sustainable strategy, with minimal detrimental effects on economy or social life. Our study strongly supports the idea that collectively abiding by strict social distancing rules can efficiently limit epidemic growth, even with full mobility, and even if these rules had not been strictly followed to the present day. The optimal course suggested by our model towards accomplishing epidemic control without the huge economic and societal burden of an extended lockdown is that of maintaining strict social distancing regulations until a clinical treatment or prevention plan is in place.

\subsection{Limitations and further work}

The model is intrinsically limited by attempting to capture social dynamics via an epidemic compartmental model. This ignores details such as household structure, and other contact patterns, which are crucial in the spread of a virus (e.g., one infected member of a household is likely to infect all other members). One line of our future work is directed towards studying similar behaviors in contact networks, and matching dynamics with the mean field information that a compartmental model provides.

Other limitations come from the gaps in the clinical and epidemiological knowledge we posses about COVID 19. Both our knowledge of the epidemic parameters, and some of the characteristics of the SARS-COV2 virus are constantly changing. This non-stationarity makes it additionally difficult to capture updated snapshots, and produce testable predictions. We aimed to build our model specifically enough to apply to the clinical and social data pertinent to the current epidemic, while keeping it sufficiently robust to generate general predictions that can be easily adapted to the rapid fluctuations in data trends.

For future iterations of the model, we are working on including the time-depending aspect in the epidemic parameters, and on incorporating some of the known feedback loops that contribute to the complexity of this system. One specific goal of our future modeling will be to study the efficiency of testing and contact tracing (which permits early detection, and adequate reduction in the travel vector of  infected, and even exposed individuals). Another important aspect we are adding to the model is the feedback introduced by reaching the health care capacity at very high infection counts (which is likely to impact the recovery versus fatality rates). Finally, a significant feedback contribution comes from the psychological effects. One obvious example is ``tolerance:'' people may be inclined to loosen isolation measures in absence of an immediate epidemic threat, thus contributing to increasing this very threat.

Finally, one very important point for future study is the network aspect. Many communities already struggling with increasing infection spread are subject  to additional exposure by individuals arriving from other locations and then acting as super-spreaders in the system. This has been happening throughout the course of the infection, but is an even higher source of concern after reopening has restored mobility (including to large meetings), and travel now occurs relatively unrestricted between US states, and even internationally to some extent. One aspect of our future work is focused on extending this modeling framework to multiple coupled communities, and on aiming to understand how social distancing and  behavior patterns within and between communities can affect the spread in the  network as a whole.

\clearpage
\section*{Supplementary File: Appendix A: mobility matrix}

Our simulation considered a community with Monday-Saturday social dynamics specified by the set of mobility matrices below. On Sunday, the dynamics is simply characterized by 60\% of people in the $S$, $E$ and $I$ compartments and all age groups attending a religious/social gathering, and no additional travel to other destinations.

\begin{table}[h!]
\begin{center}
{\renewcommand{\arraystretch}{1.25}
\begin{tabular}{|l|c|c|c|c|c|c|c|}
\hline
  & {\bf Doctor} & {\bf Store} & {\bf Church} & {\bf Campus} & {\bf School} & {\bf Park} & {\bf Restaurant}\\
\hline
{\bf S/R} & 0.01 & 0.02 & 0.1 & 0 & 0.5 & 0.3 & 0 \\
{\bf L/A/P} & 0.01 & 0.02 & 0.1 & 0 & 0.5 & 0.3 & 0\\
{\bf I} & 0.2 & 0 & 0 & 0 & 0.2 & 0.1 & 0\\
{\bf D} & 0 & 0 & 0 & 0 & 0 & 0 & 0\\
\hline
\end{tabular}}
\end{center}
\vspace{-2mm}
\caption{\small {\it {\bf Mobility array for Children.} Each entry shows the fraction of the children compartment specified by the row travels each day to the location specified by the column.}}
\label{tableA1}
\end{table}


\begin{table}[h!]
\begin{center}
{\renewcommand{\arraystretch}{1.25}
\begin{tabular}{|l|c|c|c|c|c|c|c|}
\hline
  & {\bf Doctor} & {\bf Store} & {\bf Church} & {\bf Campus} & {\bf School} & {\bf Park} & {\bf Restaurant}\\
\hline
{\bf S/R} & 0.01 & 0.1 & 0.01 & 0.4 & 0.1 & 0.01 & 0.3\\
{\bf L/A/P} & 0.01 & 0.1 & 0.01 & 0.4 & 0.1 & 0.01 & 0.3\\
{\bf I} & 0.2 & 0.1 & 0 & 0.2 & 0 & 0 & 0.2\\
{\bf D} & 0 & 0 & 0 & 0 & 0 & 0 & 0\\
\hline
\end{tabular}}
\end{center}
\vspace{-2mm}
\caption{\small {\it {\bf Mobility array for Young adults.} Each entry shows the fraction of the Young adult compartment specified by the row travels each day to the location specified by the column.}}
\label{tableA2}
\end{table}


\begin{table}[h!]
\begin{center}
{\renewcommand{\arraystretch}{1.25}
\begin{tabular}{|l|c|c|c|c|c|c|c|}
\hline
  & {\bf Doctor} & {\bf Store} & {\bf Church} & {\bf Campus} & {\bf School} & {\bf Park} & {\bf Restaurant}\\
\hline
{\bf S/R} & 0.02 & 0.15 & 0.15 & 0.15 & 0.1 & 0.15 & 0.2\\
{\bf L/A/P} & 0.02 & 0.15 & 0.15 & 0.15 & 0.1 & 0.15 & 0.2\\
{\bf I} & 0.3 & 0.1 & 0.1 & 0.05 & 0.05 & 0 & 0.1\\
{\bf D} & 0 & 0 & 0 & 0 & 0 & 0 & 0\\
\hline
\end{tabular}}
\end{center}
\vspace{-2mm}
\caption{\small {\it {\bf Mobility array for Adults.} Each entry shows the fraction of the Adult compartment specified by the row travels each day to the location specified by the column.}}
\label{tableA3}
\end{table}


\begin{table}[h!]
\begin{center}
{\renewcommand{\arraystretch}{1.25}
\begin{tabular}{|l|c|c|c|c|c|c|c|}
\hline
  & {\bf Doctor} & {\bf Store} & {\bf Church} & {\bf Campus} & {\bf School} & {\bf Park} & {\bf Restaurant}\\
\hline
{\bf S/R} & 0.1 & 0.2 & 0.3 & 0.05 & 0.05 & 0.2 & 0.05\\
{\bf L/A/P} & 0.1 & 0.2 & 0.3 & 0.05 & 0.05 & 0.2 & 0.05\\
{\bf I} & 0.4 & 0.2 & 0.1 & 0 & 0 & 0 & 0\\
{\bf D} & 0 & 0 & 0 & 0 & 0 & 0 & 0\\
\hline
\end{tabular}}
\end{center}
\vspace{-2mm}
\caption{\small {\it {\bf Mobility array for the Elderly.} Each entry shows the fraction of the Elderly compartment specified by the row travels each day to the location specified by the column.}}
\label{tableA4}
\end{table}

\clearpage
\bibliographystyle{plain}
\bibliography{references1}

\end{document}